\documentclass[a4paper,12pt]{article}
\usepackage{amsmath}
\usepackage{graphicx}
\usepackage{rotating}
\usepackage{wasysym}
\DeclareGraphicsExtensions{.pdf,.png,.jpg}
\usepackage{chicago}
\usepackage{klett}
\usepackage{def_the}

\usepackage{Fig_size_A}
\usepackage{kfig}

\renewcommand{\theequation}{\mbox{\arabic{section}.\arabic{equation}}}
\renewcommand{\thefigure}{\arabic{section}.\arabic{figure}}
\renewcommand{\thetable}{\arabic{section}.\arabic{table}}
\renewcommand{\footnoterule}{\rule{14.8cm}{0.3mm}\vspace{+1.0mm}}
\renewcommand{\baselinestretch}{1.0}
\title{Hierarchical Investing}
\author{Eckhard Platen and Renata Rendek}
\begin{document}
\thispagestyle{empty} \vspace*{1.0cm}

\begin{center}
{ \LARGE \bf  Market Efficiency \vspace*{0.5cm} and Growth Optimal Portfolio}
\end{center}

\vspace*{.5cm}
\begin{center}

{\large \renewcommand{\thefootnote}{\arabic{footnote}} {\bf Eckhard
Platen}\footnote{University of Technology Sydney,
Finance Discipline Group}$^{,}$\footnote{University of Technology Sydney, School of Mathematical and Physical Sciences}$^{,}$\footnote{University of Cape Town, Department of Actuarial Science} and {\bf Renata Rendek}\footnote{University of New South Wales,
School of Mathematics and Statistics, \\ Corresponding author: Eckhard Platen (eckhard.platen@uts.edu.au)}$^{}$}

\vspace*{2.5cm}

\today

\end{center}

\begin{minipage}[t]{13cm}
The paper predicts an Efficient Market Property for the equity market, where stocks, when denominated in units of the growth optimal portfolio (GP), have zero instantaneous expected returns. Well-diversified equity portfolios are shown to approximate the GP, which explains the well-observed good performance of equally weighted portfolios. The proposed hierarchically weighted index (HWI) is shown to be an even better proxy of the GP. It sets weights equal within industrial and geographical groupings of stocks. When using the HWI as proxy of the GP the Efficient Market Property cannot be easily rejected and appears to be very robust.
\end{minipage}
\vspace*{2.0cm}

\noindent {

\vspace*{1.0cm}

}
    \newpage
\section{Introduction}\label{section.intro}

For many decades the question, whether markets are 'efficient', e.g., as described in \citeANP{Fama70} \citeyear{Fama70,Fama91,Fama98}, has been widely discussed but never conclusively answered in the literature. The current paper aims to answer this question for developed equity markets. This answer is deeply linked to the respective growth optimal portfolio (GP). We show, when using the GP (also called benchmark) as denominator, the benchmarked (denominated in units of the GP) value of any portfolio has zero instantaneous expected returns. Thus, its current benchmarked value is the best forecast of its instantaneously following benchmarked value, which we call the \textit{Efficient Market Property}.  It resembles various types of efficient market hypotheses discussed in \citeANP{Fama70} \citeyear{Fama70,Fama91,Fama98} and a subsequent rich stream of literature. We emphasise that this is a fundamental fact and holds under extremely general model assumptions. The GP and, thus, the \textit{Efficient Market Property} are unique for a given investment universe and the respective available evolving information. This property turns out to be very robust, since we do not need to involve any particular model assumptions or estimation for constructing an excellent proxy of the GP. We then use this proxy as benchmark and show for equities of developed markets that the Efficient Market Property cannot be easily rejected. 
The findings provide a new understanding of market efficiency, which exists theoretically for any reasonable market that has a GP. When the GP does not exist for a market model then its candidate explodes in finite time, which means there is economically meaningful arbitrage in such market and the model has to be dismissed. The GP is known as the expected log-utility maximizing portfolio. It was discovered in \citeN{Kelly56} and has been widely studied; see e.g. \shortciteN{MacLeanThZi11} for an edited collection of papers. It has the fascinating property that it maximizes pathwise in the long run the long-term growth rate (GR). The GR is the by the length of the observation window normalized increment of the logarithm of the portfolio value.

The key question is then how to construct, in reality, an excellent proxy of the GP. Despite decades of research on how to construct optimal portfolios for stocks, the simple, model-independent equal-weighted approach seems to do at least as well as more complicated and theoretically grounded approaches. This stylized empirical fact has been established in \shortciteN{DeMiguelGaUp09}, where it has been shown that naive diversification (equal-weighting of stocks) outperforms most known portfolio strategies. The current paper goes beyond naive diversification and makes use of readily available information capturing key economic dependencies of stocks. This information does not involve any estimation and is obtained through classification of economic activities of respective companies. Companies, which belong to the same industrial and geographical group, are exposed to similar uncertainties and their own specific uncertainty. The hierarchical groupings provided by such classification remain rather stable over time and persist in periods of extreme market moves.  In our study we use the well-established Industry Classification Benchmark (ICB); see \shortciteN{Reuters08}.

\begin{figure}[h!] 
\centering 
\includegraphics[width=15cm]{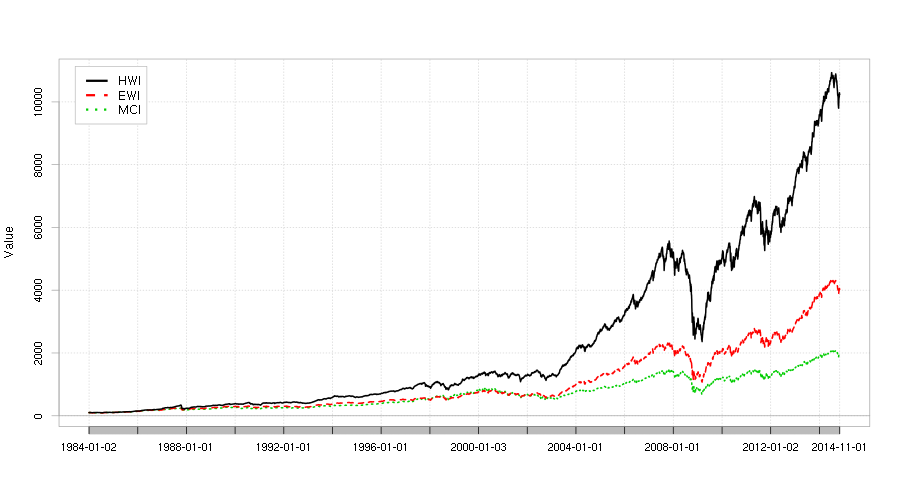}\\ 
\caption{The trajectories of the HWI, EWI and MCI.}\label{f.00} 
\end{figure}

\begin{figure}[tb] 
\centering 
\includegraphics[width=15cm]{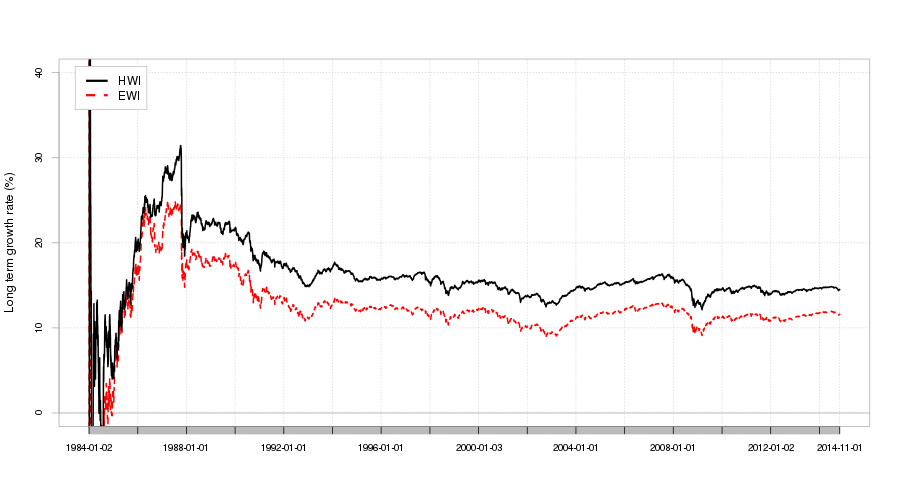}\\ 
\caption{Observed long term-growth rate (GR) (in percent) in dependence on the final observation date for the HWI and the EWI.}\label{f.0} 
\end{figure}

By naive diversification within each group and at each level of the hierarchy a new proxy for the GP is constructed, which we call the hierarchically weighted index (HWI). To illustrate the excellent performance of the HWI for stocks of the developed markets we show in Figure \ref{f.00} its trajectory together with those of the respective market capitalization weighted index (MCI) and equal-weighted index (EWI) in US dollar denomination, all starting in 1984 at an initial value of $100$. In Figure \ref{f.0} we display for the HWI and the EWI their observed (annualized) long term growth rates (GRs) in percent as functions of the end date of the observation period. For the longest available observation window, the long-term growth rate (GR) amounts for the EWI to $11.6$ and for the HWI to $14.5$, whereas the GR for the MCI reaches only $9.4$; see also Table \ref{tab.4}. Theoretically, it is the GP that obtains pathwise in the long-run the highest GR. This makes the GR for the longest available observation window a realistic performance measure that we use to distinguish between proxies of the GP.
By using the HWI as benchmark and then studying more than 30 million benchmarked stock returns we show that the theoretically predicted \textit{Efficient Market Property} cannot be easily rejected for stocks of developed markets. When choosing the MCI or EWI as benchmark, the Efficient Market Property can be easily rejected at typical significance levels.


With the HWI we propose also a new benchmark for long-term equity fund management. It is far more difficult to beat in the long run than the ‘traditional’ benchmark, the MCI. Moreover, we propose with the HWI a practically feasible approximation of a theoretically optimal portfolio, the GP. Once the GP is obtained, it is straightforward to construct other optimal portfolios with desired risk characteristics by involving the riskless asset in an appropriate manner; see e.g. Chapter 11 in Platen and Heath (2010). This allows one to overcome the main practical obstacle in portfolio optimization, identified in \citeN{DeMiguelGaUp09}.  It consist in the fact that the standard sample based mean-variance methodology of modern portfolio theory, originated by \citeN{Markowitz59}, does, in general, not gain enough accuracy to provide useful proxies for targeted optimal portfolios. The impossibility of implementing sample based portfolio optimization for large equity markets has also been pointed out, e.g., by \citeN{BestGr91}, \citeN{ChopraZi93}, \citeN{BaiNg02}, \citeN{LudvigsonNg07}, \shortciteN{PlyakhaUpVi14}, \citeN{KanZh07}, \citeN{KanWaZh16} and \citeN{OkhrinSc06}. The dilemma is that the available observation windows are too short for estimating the most likely moving expected returns. 
Building on Markowitz's mean-variance approach, see \citeN{Markowitz59}, \citeN{Sharpe64} introduced the capital asset pricing model (CAPM), which became generalized in a stream of literature. A consequence of the classical assumptions underpinning the CAPM is that the market capitalization weighted index (MCI) should maximize the Sharpe ratio. However, empirical evidence suggests that the MCI may, in reality, not yield the highest Sharpe ratio; see e.g. \citeN{HarveyLiZh14}.  This fact is also confirmed by our observations, where we report in Table \ref{tab.4} a Sharpe ratio of $0.90$ for the HWI, which is more than double that of $0.43$ for the MCI.

Since market capitalization weighted indexes seem unlikely to be mean-variance optimal, a wide range of rule-based investment strategies has emerged, aiming mostly for higher Sharpe ratios. One group of rule-based strategies uses stock characteristics, known as factors, which have been found in historical data to have an effect on the expected returns of stocks; see e.g. \citeN{RosenbergMa76}, \citeN{FamaFr92}, \citeN{Carhart97} and \shortciteN{ArnottHsMo05}. If investors systematically exploit on a large scale such factors, e.g. fundamental value, size or momentum, then most of these effects are likely to weaken or even vanish over time, as argued in \citeN{VanDijk11}. \shortciteN{HarveyLiZh14} find that some factor premia 'observed' resulted just from 'mining' the available finite data set and have to be dismissed. In summary, factor based strategies may not provide sustained high Sharpe ratios in the long run. 

Several rule-based strategies have emerged, which exploit the inverse of estimated covariance matrixes of returns. These strategies aim at, e.g. the minimum variance portfolio as in \shortciteN{ClarkeSiTh11}; the risk parity portfolio as in \shortciteN{MaillardRoTe10}; or the maximum diversification portfolio as in \citeN{ChoueifatyCo08}. As shown in \shortciteN{PlyakhaUpVi14}, even for a rather small number of stocks the arising estimation errors may easily offset the theoretical benefits of such theoretically optimal portfolio strategies.

Despite an abundance of empirical and theoretical work; see e.g. \citeN{Fernholz02}, \shortciteN{ChowHsKaLi11}, \shortciteN{LeoteLuMo12}, \shortciteN{GanderLePf13} and \citeN{Oderda15}, no authors seem to have managed to extract convincingly the theoretical reason why various rule-based strategies outperform the respective MCI. The current paper suggests such a reason through its \textit{Diversification Theorem} because it states that well-diversified portfolios approximate asymptotically the GP for increasing number of constituents and these portfolios are usually better diversified than the MCI.

The paper is organized as follows: Section \ref{section.2} summarizes facts about the GP and the Efficient Market Property. In Section \ref{section.3} we construct the HWI. Section \ref{section.4} analyzes the performance of the HWI while Section \ref{section.7} explores empirically the Efficient  Market Property. Appendix A describes the data used, whereas Appendix B proves the Efficient Market Property. Appendix C presents the Diversification Theorem. Finally, Appendix D demonstrates that the HWI equals the GP under a stylized hierarchical stock market model. 
 
\section{Efficient Market Property}\label{section.2}\setcA \setcB

In this section we summarize properties related to the \textit{growth optimal portfolio} (GP), including the Efficient Market Property.  

Let $V^\pi_t$ denote the value of a strictly positive portfolio at time $t$ with strategy $\pi$, which denotes the weights invested. We can then write the respective \textit{long-term growth rate} (GR) at time $T>0$ in the form 
\BE \label{gpi} G^\pi_T=\frac{1}{T} \ln{\Bigg{(}\frac{V^\pi_T}{V^\pi_0}\Bigg{)}}.\EE
The value of the GP, at time $t \geq 0$ is denoted by $V^{\pi^*}_t$, where we set $V^\pi_0=1$. The GP is the portfolio that maximizes the expected GR, that is, $E(G^\pi_T)$ for all $T \geq 0$. We assume that $G^{\pi^*}_T< \infty$ for all $T>0$, and have the following fundamental asymptotic relation
\BE \label{gpilim} \lim_{T \rightarrow \infty} G^\pi_T \leq \lim_{T\rightarrow \infty} G^{\pi^*}_T\EE
for any strictly positive portfolio $V^\pi$; see Theorem 3.3 in \citeN{Platen11}. 

One notes in Figure \ref{f.0} that the GRs of the HWI and the EWI fluctuate similarly, which is a consequence of the fact that both well-diversified indexes are driven mostly by the non-diversifiable uncertainty of the equity market. This observation supports our search for the 'best' proxy of the GP among various competing well-diversified portfolios by aiming for the one with the largest GR for the longest available observation window. 

As mentioned previously, the GP is also called benchmark, and values denominated in units of the GP are called benchmarked values. We will prove in Appendix B the following fundamental fact:

\begin{theorem}[Efficient Market Property]\label{t.emp}
 In a continuous market nonnegative benchmarked portfolios have zero instantaneous expected returns and zero or negative expected returns over any time period, as long as the respective GP exists.
\end{theorem}

For easier presentation, we prove this result for continuous markets, however, it holds far more generally even for semimartingale markets, see \citeN{KaratzasKa07}. The GP is in many ways the best performing portfolio. The Efficient Market Property captures a most important property where it is 'best'.
Any other portfolio, when used as benchmark, would violate for some security and some time the statements of Theorem \ref{t.emp}. The theorem says that in the very short term the current value of any nonnegative benchmarked portfolio is equal or greater than its future benchmarked values. In this precise sense the market is efficient. 

Since, in the denomination of the GP instantaneous expected returns are zero there is no possibility of maximizing expected returns. Note, only when changing the denominator, e.g. to the risk-free asset, then nonzero instantaneous expected returns appear for the stock and portfolio dynamics in the new denomination as a consequence of It\^{o} calculus. The mean-variance approach to portfolio optimization has been often aiming, at maximizing instantaneous expected returns in the denomination of the risk-free asset.
The Efficient Market Property identifies the GP as the only benchmark, where for all benchmarked portfolios their current value is over vanishing time periods the 'best' forecast of their 'next' value. The conditional expectation underpinning this statement is taken with respect to the real world probability measure and the information exploited is the one captured by the evolution of all state variables of the underlying market model up to the current time.

\section{Hierarchically Weighted Index}\label{section.3}\setcA \setcB

This section explains the construction of the \textit{hierarchically weighted index} (HWI), which we implement for stocks of the developed markets.  Details on the Industry Classification Benchmark (ICB), see \citeN{Reuters08}, which we use when forming hierarchical groupings, together with information about the data, are provided in Appendix A.

\subsubsection*{Portfolio Construction}
We assume four levels in our hierarchy for the developed stock markets. This appears to be reasonable but other numbers of levels are possible. At time $t \geq 0$ we denote by $M_t$ the number of geographical regions, $M^{j_1}_t$ the number of countries in the $j_1$-th region, $M^{j_1,j_2}_t$ the number of industrial groupings in the $j_2$-th country of the $j_1$-th region and $M^{j_1,j_2,j_3}_t$ the number of stocks in the $j_3$-th industrial grouping of the $j_2$-th country of the $j_1$-th region.

$S^j_t$ denotes the cum-dividend price of the $j$-th stock (denominated in US dollar) at time $t \geq 0$, $j=(j_1,j_2,j_3,j_4)$. The portfolio weight for the investment in the $j$-th stock at time $t$ is denoted by $\pi^j_t$. The vector process of weights of a strictly positive, self-financing portfolio with value $V^\pi_t$ at time $t \geq 0$ is (with slight abuse of notation) denoted by $\pi=\{\pi_t = (\pi^1_t, \pi^2_t,\dots, \pi^{N_t}_t)^\top, t \geq 0\}$, where  
\BE \label{Nt} N_t=\sum^{M_t}_{j_1=1} \sum^{M^{j_1}_t}_{j_2=1}\sum^{M^{j_1,j_2}_t}_{j_3=1}M^{j_1,j_2,j_3}_t\EE
denotes the number of stocks in our investment universe at time $t \geq 0$. We denote by $x^\top$ the transpose of a vector or matrix $x$.

We introduce by $0=t_0 < t_1 < \dots < t_i < t_{i+1} < \dots$ the reallocation times for a portfolio $V^\pi$. Its value $V^\pi_{t_i}$ at time $t_i$ is calculated, recursively, via the relation
\begin{equation}
V^\pi_{t_{i}}=V^\pi_{t_{i-1}}\l(1+\sum^{N_{t_{i-1}}}_{j=1}\pi^j_{t_{i-1}}\frac{S^j_{t_i}-S^j_{t_{i-1}}}{S^{j}_{t_{i-1}}}\r)
\end{equation} for $i \in \{1,2,\dots\}$ with $V^\pi_{t_0}>0$. For all portfolios constructed in this paper the rebalancing frequency is quarterly, which turns out to be adequate but can be easily changed. Additional rebalancing is performed when a stock 'dies' between the quarterly rebalancing times. More frequent rebalancing, e.g. monthly reallocation, shows similar results to what we report in our study.

The 'traditional' benchmark for fund management is the \textit{market capitalization weighted index} (MCI) with the weight $\pi^{MCI,j}_{t_i}$ invested in the $j$-th stock at time $t_i$. This weight is determined by the reported respective market value $MV^j_{t_i}$ via the formula 
\begin{equation}
 \pi^{MCI,j}_{t_i}=\frac{MV^j_{t_i}}{\sum^{N_{t_i}}_{k=1} MV^k_{t_i}}
\end{equation} for $j \in \{1,2,\dots, N_{t_i}\}$, $i \in \{0,1,\dots\}$. 

The weights $\pi^{EWI,j}_{t_i}$ for $j \in \{1,2,\dots, N_{t_i}\}$ of the \textit{equal-weighted index} (EWI) are set to be equal at time $t_i$, that is,
\begin{equation}
\label{piEWI} \pi^{EWI,j}_{t_i}=\frac{1}{N_{t_i}},
\end{equation} for $i \in \{0,1, \dots\}$. 

\subsubsection*{HWI}
\begin{figure}[tb] 
\centering 
\includegraphics[width=15cm]{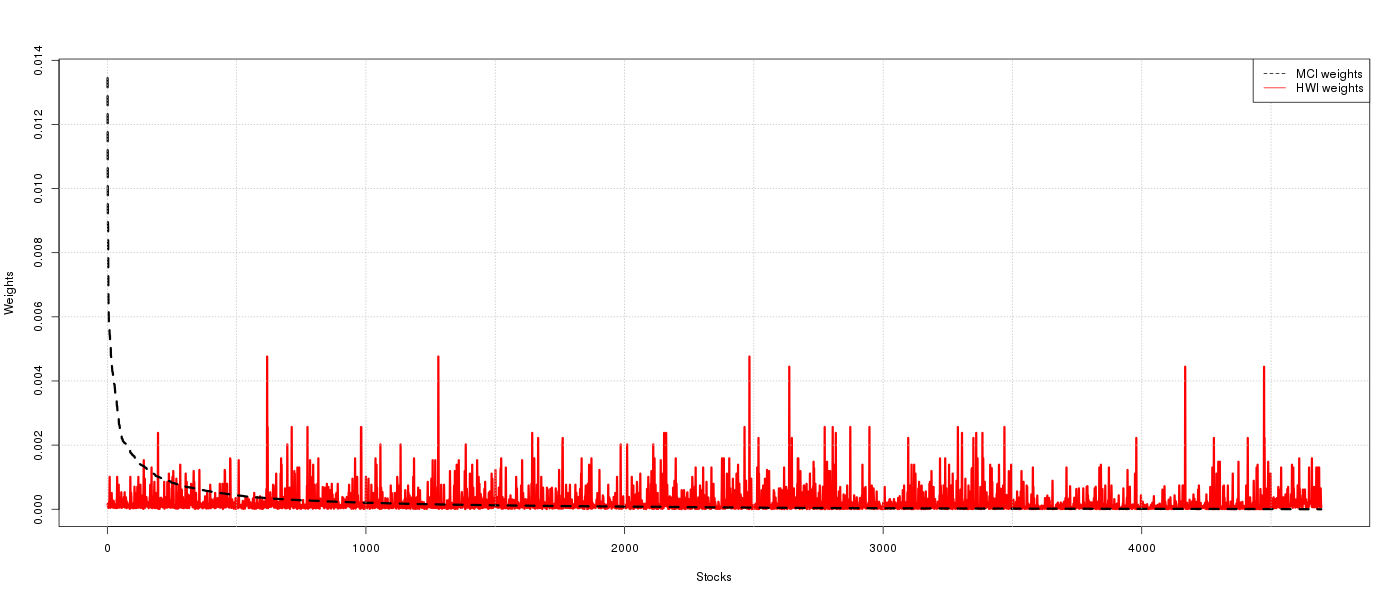}\\ 
\caption{The HWI and MCI weights of the stocks ordered by market capitalization.}\label{f.weights}
\end{figure}
The proposed \textit{hierarchically weighted index} (HWI) uses region, country and industry groupings of stocks in a hierarchical manner to determine their respective weights. It invests equal fractions of wealth in the constituents of each group. These constituents are themselves equal weighted indexes or (on the lowest level) stocks. 
The three geographical regions we distinguish are Europe, Asia-Pacific and North-America. Table \ref{tab.3}
lists in the first column the 23 developed markets (countries) considered in this
paper. These were chosen by using the ICB classification. The base
dates for investments made in each respective country are listed in column
two. These are chosen by taking into account the reported number of
dead stocks at a given time. Due to the downgrading of Greece to the status of an emerging market in 2013 we ignore Greece. We include Israel, since it became acknowledged as a developed market in 2010. 

Due to these and other deviations of our set of stocks compared to those of the MSCI Developed Market we form the MCI, which allows us to make a fairer comparison with the other constructed indexes. The constructed MCI deviates only marginally from the MSCI, as we show later on. The total number of stocks considered altogether is over $40,000$. About $4,810$ stocks are selected dynamically at a given time from these stocks by the following rule, which makes sure that there is on average, a reasonable number of stocks in the industrial groups formed within each country: The rule for the selection of subsectors, (in the ICB sense) as industrial grouping is that the country needs to have more than 900 stocks available. For countries with a number of stocks between 80 and 900 we employ the
sector grouping. For countries with less than 80 stocks the supersector grouping
is used as industrial grouping. For instance, in the case of the United States, we
choose the 998 largest (by market value) stocks that are alive at a rebalancing
date. The list of the number of stocks chosen for a given country is recorded
in column three of Table \ref{tab.3}. Furthermore, column four in Table \ref{tab.3} indicates the
level of ICB grouping we use as industrial grouping of stocks in the given country. 

Recall that, for the $j$-th stock with $j=(j_1,j_2,j_3,j_4)$ we denote by $M^{j_1,j_2,j_3}_t$ the number of stocks in the subsector, sector or supersector, respectively, according to its industrial grouping in its country. $M^{j_1,j_2}_t$ is then the
number of respective subsectors, sectors or supersectors in the country the
$j$-th stock belongs to. $M^{j_1}_t$ denotes the number of countries in the region of
the $j$-th stock and $M_t$ counts the number of regions considered at time $t \geq 0$.

We equal-weight in each group of the hierarchy the constituents we have formed at the next lower level. The weight for the $j$-th stock, with $j=(j_1,j_2,j_3,j_4)$, is then of the form
\BE \label{piHWI}\pi^{HWI,j}_t=\frac{1}{M_t}\frac{1}{M^{j_1}_t}\frac{1}{M^{j_1,j_2}_t}\frac{1}{M^{j_1,j_2,j_3}_t}\EE for $t \geq 0$. As a result, the weights of industrial groupings and countries in the HWI are rather different to those of
the MCI and EWI, see Table \ref{tw.hewi.5}. Another illustration of how different the weights of the HWI are compared to those of the MCI, is given in Figure \ref{f.weights}. It shows the weights of stocks for the MCI and the HWI, where the stocks are ordered by their market capitalization at the end of our observation period. The EWI uses no information, whereas the HWI employs information about stock groups with exposure to similar uncertainties. 

\begin{table}[h!]
\centering 
\tiny
\begin{tabular}{l|lllll}
  \hline
  \hline \\
 Country & HWI Country Base Date & No. of Stocks & Industrial Grouping\\ \\
 \hline
  \hline \\
  CANADA & 01/01/1990 &245& sector\\ 
   UNITED STATES &   02/01/1984 & 998&  subsector\\ 
   HONG KONG & 01/01/1986 & 127 &sector\\ 
  JAPAN &   01/01/1990& 1000& subsector\\ 
   UNITED KINGDOM  & 01/01/1985& 539& sector\\ 
  SPAIN & 03/01/2000& 116& sector\\ 
   NETHERLANDS &   01/01/1990& 107& sector\\ 
   AUSTRALIA &  01/01/1988&160& sector\\ 
   SWITZERLAND &  01/01/1992&146&sector\\ 
   BELGIUM &  02/01/1984&90&sector\\ 
   FRANCE &   01/01/1993& 247& sector\\ 
   GERMANY &  01/01/1990& 235& sector\\ 
   ITALY &  01/01/1986& 150& sector \\ 
   SINGAPORE &  03/01/2005&100& sector\\ 
   NORWAY &   01/01/1990& 50& supersector\\ 
   IRELAND &   01/01/1991&37& supersector\\ 
   SWEDEN &  01/01/1991&62&supersector\\ 
   FINLAND &   01/01/1996&47& supersector\\ 
   AUSTRIA &   01/01/1992&49& supersector\\ 
   PORTUGAL &   01/01/1993&48& supersector\\ 
   DENMARK &  01/01/1993&47& supersector\\ 
   NEW ZEALAND  &  03/01/2000&50& supersector\\ 
   ISRAEL &   03/01/2000&49& supersector\\ \\
   \hline
   \hline
\end{tabular}\caption{Base dates, number of stocks and type of industrial grouping used in a country for the construction of the HWI. }\label{tab.3}
\end{table}
\begin{table}[h!]
\centering
\tiny
\begin{tabular}{l|lll|l|lll}
  \hline
  \hline \\
ICB Supersector &  HWI& EWI& MCI& Country &  HWI & EWI & MCI \\  \\
  \hline
  \hline \\
Industrial Goods \& Services& 13.95 &16.19 & 11.42& CANADA & 16.67 & 5.219 &4.243 \\ 
 Personal \& Household Goods&  6.889 & 4.729&  5.272&UNITED STATES &  16.67 & 21.26& 48.51\\ 
Real Estate & 6.586 & 8.649& 3.859&HONG KONG &  6.667 & 2.706& 4.499\\ 
Technology &  6.051 &6.966 & 9.503&JAPAN & 6.667 &21.28 & 9.917 \\ 
Retail &  5.932 & 5.944& 5.492&AUSTRALIA &  6.667 & 3.409&2.819\\ 
Basic Resources &  5.917 &3.622 & 2.244&SINGAPORE &  6.667 & 2.13&1.27\\ 
  Oil \& Gas &  5.482 & 6.221& 8.978&NEW ZEALAND & 6.667 &1.065 &0.1381\\ 
 Food \& Beverage &  5.418 & 4.325& 4.587 &UNITED KINGDOM & 2.083 & 11.48 &7.865\\ 
 Financial Services &  5.146 & 7.861& 4.23 &SPAIN &  2.083 &2.45 &1.807\\ 
  Insurance &  5.033 & 2.897 & 4.137  &NETHERLANDS &  2.083 & 2.28& 1.238  \\ 
 Health Care &  5.01 & 6.242& 10.02&SWITZERLAND &  2.083 & 3.11&3.293\\ 
 Utilities&  4.662 &3.409 & 3.685 &BELGIUM &  2.083 & 1.917&0.8247\\ 
  Telecommunications &  4.491 & 1.513& 4.183&FRANCE &  2.083 & 5.262&4.531\\ 
  Travel \& Leisure &  4.393 & 4.325& 3.029 &GERMANY &  2.083 &4.985 &3.753\\ 
  Construction \& Materials &  3.624 &4.154 &1.434 &ITALY &  2.083 & 3.174&1.397 \\ 
  Media &  3.237 &2.983 & 2.6&NORWAY &  2.083 &1.065 &0.6474 \\ 
 Chemicals &  3.107 & 2.812& 2.699&IRELAND &  2.083 & 0.7882&0.1599 \\ 
  Banks &  2.956 & 4.708& 9.696&SWEDEN &  2.083 & 1.321& 1.261 \\ 
  Automobiles \& Parts &  2.111 & 2.45 & 2.934 &FINLAND &  2.083 &1.001 &0.4259\\ 
   &   & & &AUSTRIA &  2.083 &  1.044&0.2295\\ 
   &   & & &PORTUGAL &  2.083 & 1.001 &0.1515 \\ 
   &    & & &DENMARK &  2.083 &1.001 &0.6733\\ 
   &   & & &ISRAEL &  2.083 & 1.044 &0.3431\\ \\
   \hline
   \hline
\end{tabular}
\caption{Supersector and country weights for the HWI with comparison to the EWI and MCI.} 
\label{tw.hewi.5}
\end{table}

\subsubsection*{Other Hierarchical Groupings}
When forming the HWI we use the, arguably, most natural hierarchical groupings of stocks, starting with an industry grouping located in a country, which is part of a region. Hierarchically weighted indexes can also be built in other ways, e.g.,
by using only industrial or only geographical classifications of stocks. Furthermore, one may use first on the lower levels in the hierarchy both geographical and industrial groupings and at higher levels only industrial
groupings, as illustrated in \citeN{PlatenRe12an}. This would still group stocks according to exposure to similar uncertainties. 

An important question is, does it matter significantly if we go in our hierarchy on the lower level via industrial groupings or via geographical groupings? As we will see below, the performance of the resulting index appears to be better when having as lowest level an industrial grouping and on the second lowest level a geographical grouping. 

To clarify whether there may exist significantly better performing hierarchical portfolios than the proposed HWI we studied
various alternative hierarchical groupings beyond those we mention below. 
\begin{figure}[h!] 
\centering 
\includegraphics[width=15cm]{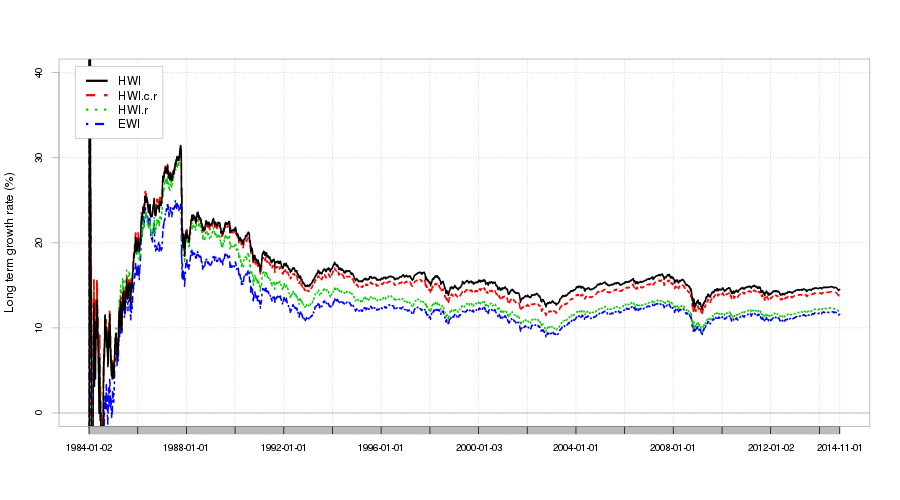}\\ 
\caption{The observed long term growth rates (GRs) (in percent) for the EWI and the hierarchically weighted indexes HWI.r and HWI.c.r, diversified by geographical origin, together with those of the HWI.}\label{fig.1} 
\end{figure} 
To illustrate our findings, we report observed long-term growth rates (GRs) for selected hierarchically weighted indexes and compare these to the HWI and EWI. First, we mention two illustrative examples of indexes that are diversified by their geographical
origin at the stock level. These are the hierarchically weighted index diversified by region
only (HWI.r) and the hierarchically weighted index diversified by country and then by region
(HWI.c.r). The GRs of the EWI and these two indexes are compared to the GR of the HWI in Figure \ref{fig.1}, where we observe that the HWI outperforms the HWI.c.r, which outperforms the HWI.r, and the latter the EWI.
This figure illustrates also that the addition of an extra hierarchical level provides consistently an improvement in the GR. Note that in Figure \ref{fig.1} the use of information about the region provides already a slight improvement over the performance of the
EWI. Introducing the country level in the hierarchy produces visually in the long run in Figure \ref{fig.1} the main
improvement. The GR improves further by introducing appropriate industrial groupings of stocks in each country, which leads to the HWI. 

We emphasize that the GRs of all these indexes fluctuate very similarly, which allows us to distinguish already after a few years of observation between their GRs. The index with the highest GR at the end of the observation period is the HWI, as is evidenced also by Table \ref{tab.4p} in the next section.

For further illustration, we mention three other examples of indexes grouped by their
industrial origin on the stock level. These are: the hierarchically weighted index HWI.s by global industry sector only, the hierarchically weighted index HWI.c.g diversified by country industry
sector first and global industry sector second and the hierarchically weighted index HWI.c.r.g, diversified first by country sector, then by regional sector and, finally, by global sector. In comparison to these indexes and all similar indexes studied we find that the proposed HWI generates the highest observed GR. For this reason, and by the theoretical underpinning we give in Appendix C and Appendix D, we consider the HWI as the best proxy of the GP for the stocks of developed markets among the indexes studied in this paper.

\section{Further Empirical Results}\label{section.4}\setcA \setcB

This section presents further empirical results concerning the performance of the HWI, EWI and MCI. Table \ref{tab.4} reports in column two (in percent) the observed (annualized) long-term growth rate (GR) over the available observation window of $T=31$ years, calculated according to formula (\ref{gpi}). We emphasize that this has been our key performance measure throughout the paper, since it targets directly the GP when maximized. The traditionally in equity fund management used benchmark, the MSCI total return index for the developed markets, is additionally included in Tables \ref{tab.4}-\ref{tab.dr} for comparison. This index draws stocks from
the same 23 developed markets we consider. However, it is based only on approximately 1,700
stocks, while the HWI, the MCI and our other indexes are based on over 4,700 stocks. Since the investment universe of the MSCI captures less sources of uncertainty than that of the MCI, one should expect a lower GR for the GP of the constituents of the first one, which most likely leads also to the lower GR observed for the MSCI compared to the one of the MCI.

\begin{figure}[h!] 
\centering 
\includegraphics[width=15cm]{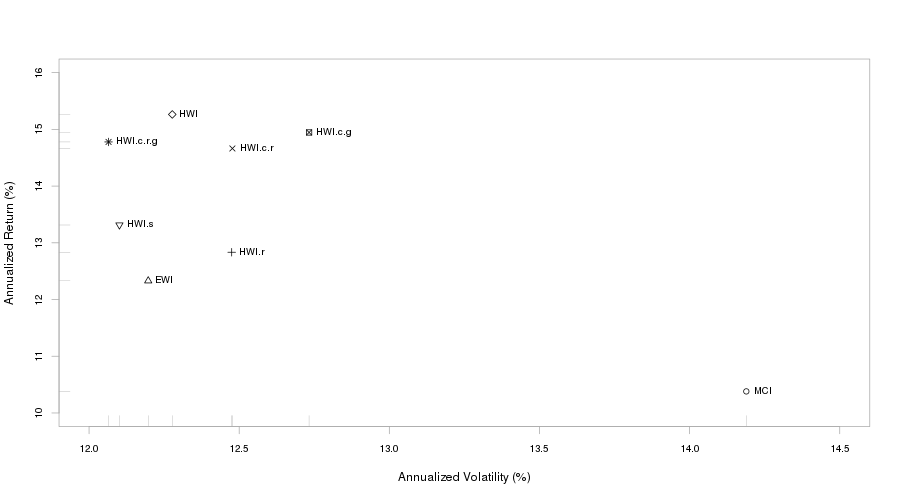}\\ 
\caption{Average return versus volatility of selected indexes (in percent).}\label{fig.2} 
\end{figure} 
 \begin{table}[h!]
\centering
\tiny
\begin{tabular}{l|rrrrrrr}
  \hline
  \hline \\
Index & GR & Average Return & Risk Premium & Volatility & Sharpe Ratio  \\ \\
  \hline
  \hline \\
MSCI & 9.23 & 10.48 & 6.26 & 15.79 & 0.3963  \\ 
  MCI & 9.37 &10.38 & 6.16 & 14.19 &  0.4343  \\ 
  EWI & 11.58 &12.33 & 8.11 & 12.20 &  0.6650  \\ 
  HWI & 14.50 &15.26 & 11.05 & 12.28 &  0.8997 \\ \\
   \hline \hline
\end{tabular}
\caption{Long-term growth rate (GR), average return, risk premium, volatility and Sharpe ratio for the MSCI, MCI, EWI and HWI.}\label{tab.4}
\end{table}

\begin{table}[h!]
\centering
\tiny
\begin{tabular}{l|rrrrr}
  \hline \hline \\
Period (years) & MSCI & MCI & EWI & HWI.r & HWI.c.r \\ \\
  \hline \hline \\
1 & 5.418 (5.220,5.616) & 5.216 (5.040,5.392) & 2.783 (2.680,2.885) & 2.536 (2.438,2.635) & 0.6297 (0.5989,0.6606) \\ 
  2 & 5.518 (5.364,5.673) & 5.350 (5.223,5.477) & 2.938 (2.877,2.999) & 2.633 (2.573,2.693) & 0.6329 (0.6080,0.6578) \\ 
  3 & 5.631 (5.494,5.769) & 5.351 (5.240,5.461) & 3.005 (2.963,3.047) & 2.654 (2.611,2.697) & 0.6290 (0.6066,0.6514) \\ 
  4 & 5.771 (5.642,5.900) & 5.323 (5.221,5.425) & 2.989 (2.957,3.022) & 2.728 (2.695,2.761) & 0.6312 (0.6102,0.6523) \\ 
  5 & 5.914 (5.793,6.035) & 5.303 (5.208,5.399) & 3.012 (2.986,3.037) & 2.811 (2.786,2.836) & 0.6413 (0.6212,0.6614) \\ 
  6 & 5.973 (5.859,6.086) & 5.248 (5.160,5.336) & 3.002 (2.979,3.025) & 2.847 (2.825,2.869) & 0.6504 (0.6312,0.6696) \\ 
  7 & 5.964 (5.856,6.073) & 5.172 (5.090,5.254) & 2.990 (2.968,3.012) & 2.870 (2.849,2.891) & 0.6639 (0.6452,0.6826) \\ 
  8 & 5.986 (5.880,6.091) & 5.154 (5.076,5.231) & 2.992 (2.972,3.011) & 2.904 (2.886,2.921) & 0.6696 (0.6515,0.6877) \\ \\
   \hline \hline
\end{tabular}
\caption{Difference in average percentage long term growth rate (GR) over observation windows reaching from one to eight years  with 95\% confidence interval between HWI and MSCI, MCI, EWI, HWI.r and HWI.c.r, respectively.}\label{tab.4p}
\end{table}

\begin{table}[h!]
\centering
\tiny
\begin{tabular}{l|rr}
  \hline
  \hline \\
 & EWI & HWI \\ \\
  \hline
  \hline \\
Daily & 0.526793 & 0.539724 \\ 
  Monthly & 0.557688 & 0.605034 \\ 
  Quarterly & 0.596067 & 0.654559 \\ 
  Half-yearly & 0.618356 & 0.689812 \\ 
  Yearly & 0.670045 & 0.747530 \\ 
  2 Yearly & 0.748774 & 0.866499 \\ 
  3 Yearly & 0.801810 & 0.885338 \\ 
  5 Yearly & 0.777516 & 0.886695 \\ \\
   \hline
   \hline
\end{tabular}
\caption{Relative frequency of outperforming the MCI over a given period length for the EWI and HWI.}\label{tab.prob}
\end{table}
\begin{table}[h!]
\centering
\tiny
\begin{tabular}{l|rr}
  \hline
  \hline \\
 Index& VaR (95\%) & ES (95\%)  \\ \\
  \hline
  \hline \\ 
MCI & -0.012834 & -0.021029 \\ 
  EWI & -0.011252 & -0.018382 \\ 
  HWI & -0.010636 & -0.018680  \\  \\
   \hline
   \hline
\end{tabular}
\caption{Value at Risk (VaR) and Expected Shortfall (ES) for MCI, EWI and HWI at a $95\%$ level.}\label{tab.var}
\end{table}
\begin{table}[h!]
\centering
\tiny
\begin{tabular}{l|rr}
  \hline \hline \\
 Index & Av. Drawdown & Av. Recovery  \\ \\
  \hline \hline \\
  MCI &  0.0199& 14.7533 \\ 
  EWI & 0.0187&  11.9556\\ 
  HWI & 0.0169 &  9.4541\\ \\
   \hline \hline 
\end{tabular}
\caption{Average relative drawdown and recovery time (in days) for the MCI, EWI and HWI.}\label{tab.dr}
\end{table}

We show in Table \ref{tab.4p} the average (with 95\% confidence intervals) of the difference between the GRs of the HWI and those of the MSCI, MCI, EWI, HWI.r and HWI.c.r, respectively, estimated from all available observation windows of length 1,2,3,4,5,6,7 and 8 years. Note that the HWI performs for all observation windows best. As we have seen in Figure \ref{f.00} and observe now in Table \ref{tab.4p}, due to the similar fluctuations of the considered well-diversified portfolios it seems to take only an observation window of about one year to distinguish reasonably well between the GRs of the HWI, the MCI and the EWI. We emphasize that in Table \ref{tab.4p} one can observe with $95\%$ confidence that already after one year the typical difference between the GRs of the HWI and MCI is at least about $5\%$. This is a substantial difference in performance, which only decreases by about $0.35\%$ when taking realistic $40$ basis points proportional transaction costs into account, as shown below in Table \ref{tab.4.tc}.

Despite our strategic focus on maximizing the GR we provide in this section also some popular short term performance and risk measures. The annualized percentage average returns are displayed in column three of Table \ref{tab.4} and
the risk premium is estimated in column four. The risk premium of the HWI is
the highest and reaches approximately 11\% compared to 6\% for the MCI. The annualized percentage volatility is recorded
in column five, and equals 12\% for the HWI, which is close to the volatility of
the EWI, whereas the volatility of the MCI is higher with about 14\%.

Figure \ref{fig.2} plots the annualized average daily return versus the annualized
volatility for selected indexes, including those with alternative hierarchical groupings mentioned in the previous section. The HWI exhibits the most favorable annualized average
return. Its Sharpe ratio is the highest, as shown in column six of Table
\ref{tab.4}. Furthermore, in Table \ref{tab.4} we see that the improvement of the HWI in its 
average return above that of the MCI is about 4.9\%, which is above the one of the EWI, which reaches about 2.9\%. As already indicated in the introduction, the Sharpe ratio of the MCI is about 0.43, the one of the EWI equals 0.67
and that of the HWI amounts to about 0.90. The latter is the highest Sharpe ratio observed in our study. One has to conclude that the MCI is unlikely positioned at the respective mean-variance efficient frontier. As discussed earlier, this observation is defying empirically classical theory, as developed in \citeN{Markowitz59}, \citeN{Sharpe64} and a related stream of literature.

Table \ref{tab.prob} provides us with observed relative frequencies for outperforming
the MCI for a given period length. Note, the largest relative frequencies are observed
for the HWI. In Table \ref{tab.var} the HWI shows the smallest
absolute values for daily Value at Risk (VaR) and Expected Shortfall (ES), respectively, on a 95\% quantile
level compared to those of the MCI and the EWI.

In Table \ref{tab.dr} we summarize for the different indexes the average drawdown
relative to the running maximum and the average time in days of recovery back to
the level of the running maximum. Again, the HWI performs best when compared
to the MCI and the EWI. This is consistent with a theoretical prediction in
\citeN{KardarasPl10me}, where the GP is shown to need the shortest expected 'market
time' to reach a target level. 

 \begin{table}[h!]
\centering
\tiny
\begin{tabular}{l|rrrrrrr}
  \hline
  \hline \\
Index & GR &Average Return & Risk Premium & Volatility & Sharpe Ratio \\ \\
  \hline
  \hline \\
  MCI-TC & 9.20 &10.22 & 6.00 & 14.19 &  0.4228 \\ 
  EWI-TC & 11.30  &12.05 & 7.83 & 12.20 &  0.6423  \\ 
  HWI-TC & 14.15 &14.92 & 10.70 & 12.28 &  0.8714 \\ \\
   \hline \hline
\end{tabular}
\caption{Long term growth rate (GR) and other common statistics for the HWI-TC, EWI-TC, MCI-TC constructed with 40 basis points proportional transaction costs.}\label{tab.4.tc}
\end{table}

From an equity fund management perspective one needs to ask about the impact of transaction costs. In Table \ref{tab.4.tc} we are imposing 40 basis points proportional transaction costs on every transaction, leading from the HWI, EWI and MCI to the HWI-TC, EWI-TC and MCI-TC, respectively. The turnover for the HWI is surprisingly low. Table \ref{tab.4.tc} shows only minor changes in performance estimates that have to be compared with the respective values in Table \ref{tab.4}. Most important is that we observe only a minor decrease of the GR for the HWI from $14.50\%$ shown in Table \ref{tab.4} to the GR of the HWI-TC of $14.15\%$ shown in Table \ref{tab.4.tc}. This makes the HWI-TC a valuable long-term investment security that can be efficiently implemented in practice. 

\section{Efficient Market Property} \label{section.7}\setcA \setcB

We are now ready to employ the constructed indexes for attempting to demonstrate empirically that the Efficient Market Property cannot be easily rejected. As shown in Theorem \ref{t.emp}, a crucial property of the GP is that when used as benchmark it makes the expected returns of nonnegative benchmarked portfolios negative or zero but never strictly positive. Thus, one can reject any candidate proxy for the GP by showing at a respective significance level that the mean of returns of benchmarked securities is strictly positive. To attempt this for any single stock would not work, since the available observation window, here 31 years, is clearly too short to provide any reasonable level of significance. However, we can gain sufficient evidence by combining all available daily returns of all benchmarked stocks in a large sample of $31,472,596$ daily returns, where we employ the HWI-TC as benchmark. For comparison, we employ also the MCI-TC, EWI-TC, HWI.c.r-TC, HWI.c.g-TC and HWI.c.r.g-TC as benchmarks to see, whether the theoretical Efficient Market Property, in the sense of Theorem \ref{t.emp}, can be potentially not rejected for several of these indices.

\begin{table}[h!]
\centering
\tiny
\begin{tabular}{l|rrrrrr}
  \hline
  \hline \\
Benchmark & Sample mean & Standard Error & 99\% LCI & 99\% UCI & Z-test & p-value \\ \\
  \hline
  \hline \\
  MCI-TC & 3.504079 & 0.142278 & 3.137594 & 3.870563 & 24.628 & 0 \\
  EWI-TC & 0.936921 & 0.141376 & 0.572761 & 1.301081 & 6.627 & 0 \\
  HWI-TC & -1.671584 & 0.141828 &-2.036909 &-1.306259 &-11.786 & 1 \\
  HWI.c.r-TC & -1.072318 & 0.141815 &-1.437608& -0.707027 & -7.561 &  1 \\ 
  HWI.c.g-TC & -1.238168 & 0.141928 & -1.603750 & -0.872587 & -8.724 &  1\\
  HWI.c.r.g-TC &-1.364769 &  0.141681 & -1.729714& -0.999823 & -9.633 &  1 \\ \\
   \hline \hline
\end{tabular}
\caption{Test for the mean daily annualized percentage returns of all benchmarked stocks.}\label{t.7}
\end{table}

\begin{table}[h!]
\centering
\tiny
\begin{tabular}{l|rrrrrr}
  \hline
  \hline \\
Benchmark & Bootstrap mean &  99\% LCI & 99\% UCI & Test statistic & p-value \\ \\
  \hline
  \hline \\
  MCI-TC & 3.502956 & 3.140459 & 3.910323 & 23.133 & 0 \\
  EWI-TC & 0.936228 & 0.571123 & 1.398584 & 6.256 & 0 \\
  HWI-TC & -1.664403 & -2.018709 & -1.277940 & -11.529 & 1\\
  HWI.c.r-TC & -1.074673 & -1.419532 & -0.674604 & -7.392& 1\\ 
  HWI.c.g-TC & -1.243531 & -1.617380 & -0.893065 &  -8.177 & 1 \\
  HWI.c.r.g-TC &  -1.361862 & -1.725689 & -1.001110 & -8.968 & 1\\ \\
   \hline \hline
\end{tabular}
\caption{Bootstrap test for the mean daily annualized percentage returns of benchmarked stocks.}\label{t.7p}
\end{table}


 We consider all through the data available daily, annualized percentage returns of all stocks (that constitute the HWI) when benchmarked by the index shown in column one of Table \ref{t.7}, producing the respective sample mean displayed in column two of Table \ref{t.7}. In a first step one could argue that the considered returns are reasonably independent when observed at different days, and that the returns of different benchmarked stocks at the same day are also reasonably independent because they are mainly driven by their idiosyncratic or specific uncertainties. Therefore, assuming in the first step of our analysis independent and identically distributed returns, appears to be acceptable. The Central Limit Theorem determines then the length of the respective confidence intervals. In Table \ref{t.7} we show in column three the resulting standard error and in columns four and five the lower level (LCI) and the upper level (UCI), respectively, of the $99\%$ confidence interval for the 'true' expected daily return of benchmarked stocks. The reader may be surprised to see some confidence intervals only covering negative values. However, this is what Theorem \ref{t.emp} predicts when considering returns over some positive time period, here about one day. Theorem \ref{t.emp} explains this phenomenon via the, so called, supermartingale property of benchmarked securities, see also \citeN{PlatenHe10} or \citeN{KaratzasKa07} for further details.
 
 To be precise, we denote (in line with Theorem \ref{t.emp} and (\ref{b.10p})), by $\mu$ the 'true' expected daily return of benchmarked stocks and test the hypothesis:
\BE \label{mu1}H_0: \mu \leq 0 \ \ \ \  \text{   versus } \ \ \  H_1: \mu>0.\EE We display the corresponding test statistic of the well-known Z-test, (see e.g. \citeN{Mode66}), in column six and the respective one-sided p-values in column seven of Table \ref{t.7}. On a $1\%$ level of significance we can clearly reject $H_0$ for the MCI-TC and EWI-TC when used as benchmarks. However, we cannot reject $H_0$ for the HWI-TC, HWI.c.r-TC, HWI.c.g-TC and HWI.c.r.g-TC. This means, these hierarchically constructed indexes seem to be closer to the GP than the MCI-TC and EWI-TC.

One may argue that the assumption on independent and identically distributed returns in the first step of our analysis is too strong and should be relaxed. Therefore, in the second step of our study we remove this assumption and report in Table \ref{t.7p} the block bootstrap percentile $99\%$ confidence intervals, with the respective test statistics and p-values for the hypothesis (\ref{mu1})\footnote{The block bootstrap replicates of the sample mean are obtained with the tsboot() function in the boot package in R, see e.g. \citeN{DavisonHi07}, using block resampling with block lengths having a geometric distribution.}. We note that the results in Table \ref{t.7p} resemble those in Table \ref{t.7}.

We emphasize in Table \ref{t.7} and Table \ref{t.7p} for the not rejected indexes that the $99\%$ confidence intervals for the mean $\mu$ do, in all cases, not include zero and cover slightly negative values, a consequence of the theoretically predicted supermartingale property of the by the GP benchmarked stocks. We note that when using the HWI-TC as proxy of the GP, the $99\%$ confidence interval for the mean of the daily returns of benchmarked stocks turns out to be the most 'negative' one, which supports our choice of the HWI-TC as the best proxy of the GP among the considered indexes. 


\begin{table}[h!]
\centering
\tiny
\begin{tabular}{l|rrrrr}
  \hline
  \hline \\
Benchmark & Sample mean &	99 \% LCI 	& 99 \% UCI &	Z-test& 	p-value \\ \\
\hline \hline \\
MCI-TC&-5.239304	&-8.707531&	-1.771077&	-3.89&	1\\
EWI-TC&-2.943474	&-4.914147&	-0.972799	&-3.85&	1\\
HWI.c.r-TC&-0.587549&	-1.219394&	0.044296&	-2.4&	0.99\\
HWI.c.g-TC&-0.440414&	-2.217511&	1.336683&	-0.64&	0.74\\
HWI.c.r.g-TC&-0.477921&	-1.354879&	0.399037&	-1.4&	0.92\\ \\
   \hline \hline
\end{tabular}
\caption{Test for the mean daily annualized returns of selected, with the HWI-TC benchmarked portfolios. }\label{t.8p}
\end{table}


We noted already that well-diversified candidate proxies of the GP are driven by similar uncertainties and can be compared empirically with each other, even over relatively short observation windows. In Table \ref{t.8p} we test the average daily returns of the with the HWI-TC benchmarked MCI-TC, EWI-TC, HWI.c.r-TC, HWI.c.g-TC and HWI.c.r.g-TC according to the hypothesis (\ref{mu1}). We note that the hypothesis $H_0$ in (\ref{mu1}) cannot be rejected for the HWI-TC when used as benchmark also for each of the above indexes. Since all the p-values are above $0.7$ or much larger, also here we cannot reject the Efficient Market Property at the $1\%$ level of significance. We repeated this study using the previously applied block bootstrap methodology and obtained very similar results.

Our deliberately simple designed empirical study indicates that it is difficult to reject the theoretically predicted Efficient Market Property for stocks of the developed market by employing the HWI-TC as proxy of the respective GP. In \citeANP{Fama70} \citeyear{Fama70,Fama91,Fama98} and subsequent literature various efficient market hypotheses have been proposed and empirically studied. Important for these forms of market efficiency is the degree of information exploited. We have now seen that such information does not seem to be highly relevant. We only need to take into account information about the natural industrial and geographical grouping of stocks to construct the HWI-TC, our best proxy of the GP for which the Efficient Market Property has not been rejected. This means, due to the well-diversified nature of the GP for stocks of the developed markets the Efficient Market Property of this market is rather robust.

It is beyond the purpose of this paper to go any further into the direction of empirically studying the Efficient Market Property. Our aim is here to provide a new understanding of the objectively given market efficiency and to open a new direction for empirical research. 

\section{Conclusion}
 The paper reveals a deep connection between the Efficient Market Property and the growth optimal portfolio (GP), since stocks denominated in units of the GP exhibit zero or negative expected returns.  Due to the impossibility of estimating means and covariances of stock returns, theoretically optimal stock portfolios, including the GP, cannot be implemented accurately enough for larger stock markets to be useful. The paper approximates the GP for the stocks of developed markets by a hierarchically weighted index (HWI), which does not rely on any estimation and sets weights equal within industrial and geographical groupings of stocks. A Diversification Theorem explains why naive diversification works rather well and the HWI performs even better. The HWI is the stock portfolio with the highest observed long-term growth rate among the considered well-diversified portfolios. The Efficient Market Property turns out to be difficult to reject empirically when using the HWI as proxy of the GP. Since no information is needed beyond that encapsulated in prices of stocks and their industrial and geographical hierarchical groupings, the Efficient Market Property appears to be very robust. These findings provide a new understanding of market efficiency, which appears to be always present as long as the GP exists for the given market. They show that by constructing an excellent proxy of the GP for a particular market one can empirically test for market efficiency. What is needed for such a test is a sufficiently accurate proxy of the GP, which is an optimal portfolio that gives access to the construction of many other optimal portfolios.  The HWI, as proxy of the GP for the stocks of developed markets, is in itself useful in equity fund management and can serve as building block in portfolio optimization and risk management. The GP plays a central role as num\'{e}raire  portfolio for pricing and hedging under the real world probability measure in the benchmark pricing theory, which goes significantly beyond classical finance with its richer modeling world and coverage of new phenomena. The HWI, as excellent proxy of the GP, allows one to practically demonstrate and exploit such new phenomena that can potentially explain various ‘puzzles’ in classical finance as forthcoming work will demonstrate.  

\appendix

\textwidth14.5cm \textheight9in
\topmargin0pt
\renewcommand{\theequation}{\mbox{A.\arabic{equation}}}
\renewcommand{\thefigure}{A.\arabic{figure}}
\renewcommand{\thetable}{A.\arabic{table}}
\renewcommand{\footnoterule}{\rule{14.8cm}{0.3mm}\vspace{+1.0mm}}
\renewcommand{\baselinestretch}{1.0}
\pagestyle{plain}

\section*{Appendix A: Data}\setcA \setcB
In this appendix we describe the data employed. Thomson Reuters Datastream (TRD) provides a range of TRD calculated country, region and sector indexes together with their current constituents and lists of dead stocks. For the
developed stock market this paper builds global stock indexes from the data available in the TRD database.
The 23 developed countries, included in this study, are given in the first column of
Table \ref{tab.1}. These developed countries are identified based on the FTSE/ICB country
classification. For each of these developed markets Datastream uses a sample of
stocks covering a minimum of 75 - 80\% of total market capitalization by choosing the largest stocks by market value. Table \ref{tab.1} lists in column two for the selected 23 countries the corresponding approximate number of stocks in the Datastream index and in column three the relevant base date from which the index is available; see
also \citeN{Reuters08}.

The respective list of constituents of Datastream country indexes has been obtained from the TRD database by quoting the mnemonic for each country list.
Table \ref{tab.2} lists the mnemonics used for the active and dead stocks
in the considered markets. We also provide in this table the number of active
and dead stocks present on both mentioned lists for each country. Note that we
only consider those stocks in the lists whose "GEOLN"="GEOGN"="Country
Name". Additionally, companies with datatype "MAJOR"="Y" are included,
which means that for companies with more than one equity security the one with
the largest market capitalization is chosen.
\begin{table}[h!]
 \centering
\tiny{\begin{tabular}{llll}
  \hline
  \hline \\
 Country & Approx. no. of stocks& Base Date  \\ \\
  \hline \hline \\
CANADA & 250 & Jan 1973  \\ 
UNITED STATES & 1000 & Jan 1973  \\ 
HONG KONG & 130 & Jan 1973  \\ 
JAPAN & 1000 & Jan 1973 \\ 
UNITED KINGDOM & 550 & Jan 1965 \\ 
SPAIN & 120 & Jan 1986  \\ 
NETHERLANDS & 130 & Jan 1973 \\ 
AUSTRALIA & 160 & Jan 1973 \\ 
SWITZERLAND & 150 & Jan 1973 \\ 
BELGIUM & 90 & Jan 1973  \\ 
FRANCE & 250 & Jan 1973 \\ 
GERMANY & 250 & Jan 1973  \\ 
ITALY & 160 & Jan 1973  \\ 
 SINGAPORE & 100 & Jan 1973  \\ 
 NORWAY & 50 & Jan 1980  \\ 
 IRELAND & 50 & Jan 1973 \\ 
 SWEDEN & 70 & Jan 1982 \\ 
 FINLAND & 50 & Mar 1988 \\ 
 AUSTRIA & 50 & Jan 1973 \\ 
 PORTUGAL & 50 & Jan 1990 \\ 
 DENMARK & 50 & Jan 1973  \\ 
  NEW ZEALAND & 50 & Jan 1988 \\ 
 ISRAEL & 50 & Jan 1992 \\ 
   \hline \hline 
\end{tabular}}\caption{Base dates and number of constituents for Datastream indexes}\label{tab.1}
\end{table}
\begin{table}[h!]
 \centering
\tiny{\begin{tabular}{l|lrrlrrr}
  \hline
  \hline \\
  Country & Active& No. Active & Downl.  & Dead& No. Dead& Downl. & Removed. \\ \\
  \hline \hline \\
CANADA & LTOTMKCN & 250 & 245 &  DEADCN1-2 & 6814 & 5310 & 598 \\ 
  UNITED STATES & LTOTMKUS & 999 & 998  & DEADUS1-6 & 22189 & 17009& 467\\ 
  HONG KONG & LTOTMKHK & 130 & 127  & DEADHK & 248 & 200  & 8\\ 
  JAPAN & LTOTMKJP & 1000 & 1000 & DEADJP & 1681 & 1569  & 14\\ 
  UNITED KINGDOM & LTOTMKUK & 549 & 539 & DEADUK & 5625 & 5263& 1037\\ 
  SPAIN & LTOTMKES & 120 & 116 &  DEADES & 264 & 180  &9\\ 
 NETHERLANDS & LTOTMKNL & 117 & 107  & DEADNL & 429 & 343 & 39\\ 
  AUSTRALIA & LTOTMKAU & 160 & 160  & DEADAU & 1784 & 1531 & 32\\ 
   SWITZERLAND & LTOTMKSW & 150 & 146  & DEADSW & 360 & 246 & 12\\ 
   BELGIUM & LTOTMKBG &  90 &  90  & DEADBG & 271 & 245  &23\\ 
   FRANCE & LTOTMKFR & 250 & 247  & DEADFR & 1534 & 1400  &243\\ 
  GERMANY & LTOTMKBD & 250 & 235  & DEADBD& 3000 & 2229 &21 \\ 
  ITALY & LTOTMKIT & 160 & 150  & DEADIT & 422 & 339  &24\\ 
  SINGAPORE & LTOTMKSG & 100 & 100  & DEADSG & 409 & 391  &4\\ 
  NORWAY & LTOTMKNW &  50 &  50  & DEADNW & 415 & 400  &34\\ 
  IRELAND & LTOTMKIR &  37 &  37  & DEADIR & 129 & 108  &22\\ 
  SWEDEN & LTOTMKSD &  70 &  62  & DEADSD & 819 & 709  &72\\ 
  FINLAND & LTOTMKFN &  50 &  47  & DEADFN & 149 & 124  &17\\ 
  AUSTRIA & LTOTMKOE &  50 &  49  & DEADOE & 196 & 160  &8\\ 
  PORTUGAL & LTOTMKPT &  50 &  48  & DEADPT & 239 & 165  &51\\ 
  DENMARK & LTOTMKDK &  50 &  47  & DEADDK & 277 & 254  &15\\ 
  NEW ZEALAND & LTOTMKNZ &  50 &  50  & DEADNZ & 252 & 224  &4\\ 
  ISRAEL & LTOTMKIS &  50 &  49  & DEADIS & 505 & 413  &1\\ \\
   \hline
   \hline 
\end{tabular}}\caption{Datastream stock lists and number of equity data obtained}\label{tab.2}
\end{table}

TRD classifies equities according to the previously mentioned Industry Classification Benchmark (ICB). We only consider those stocks that are classified into one of the subsectors and remove
from our sample all stocks that are unclassified (UNCLAS) or classified as one
of the following: unqoted equities (UQEQS), exchange traded funds (NEINV),
suspended equities (SUSEQ) and other equities (OTHEQ). The number of such
removed securities is recorded in the last column of Table \ref{tab.2}. 

We use the ICB classification of stocks into subsectors, sectors and super-
sectors, downloaded from Thomson Reuters Datastream (TRD) with mnemonics
FTAG3, FTAG4, FTAG5. The country where the given stock originates is also
recorded by TRD. Finally, the countries are grouped into three regions: Americas,
EMEA and Asia-Pacific.

The number of downloaded stocks for each active and dead list is recorded
in Table \ref{tab.2} in the fourth and seventh column, respectively. For the downloaded stocks we
obtained the total return prices and the market capitalization. The market value
is here understood as the reported number of ordinary shares in the market
multiplied by the stock price. Note that we have accounted for the fact that TRD
repeats the last valid stock price or market capitalization for delisted stocks after a
delisting. We find it necessary to remove this zero return from
the end of the time-series. With the downloaded stock data prepared in this manner we perform our study, where we recover
well the by TRD historically formed market capitalization weighted indexes for the 23 developed markets. We
then use this data for constructing the EWI, MCI and hierarchically weighted
indexes previously mentioned.

\newtheorem{corollary}[theorem]{Corollary}
\newtheorem{assumption}[theorem]{Assumption}

\renewcommand{\theequation}{\mbox{B.\arabic{equation}}}
\renewcommand{\thefigure}{B.\arabic{figure}}
\renewcommand{\thetable}{B.\arabic{table}}
\renewcommand{\thesubsubsection}{B.\arabic{section}}
\renewcommand{\thetheorem}{B.\arabic{theorem}}
\renewcommand{\thecorollary}{B.\arabic{theorem}}
\renewcommand{\theassumption}{B.\arabic{theorem}}

\section*{Appendix B: Efficient Market Property}\setcA \setcB
\subsection*{Market Setting}
We prove in this appendix several theoretical results, which underpin our reasoning when deriving the Efficient Market Property and constructing well-diversified portfolios to approximate the GP. To avoid technicalities, we consider in our proofs and derivations a continuous financial market. However, the Efficient Market Property can be shown to hold for general semimartingale markets, see \citeN{KaratzasKa07} and \citeN{DuPlaten16}. The traded uncertainty of stocks is modeled by an $n$-dimensional standard Brownian motion $W=\{W_t=(W^1_t, \dots,W^n_t)^\top, t \in [0,\infty)\}$, $n \in \{2,3,\dots\}$ on a filtered probability space $(\Omega, \mathcal{F},\underline{\cal{F}},\\ P)$, satisfying the usual conditions, see e.g. \citeN{KaratzasSh98}, where the filtration $\underline{\cal{F}}$ $=(\mathcal{F}_t)_{t \geq 0}$ models the evolution of information, generated by the quantities constituting the model. This information is characterized at time $t \geq 0$ by the sigma algebra $\mathcal{F}_t$, satisfying the usual conditions, see \citeN{KaratzasSh98}. Concerning the Efficient Market Property information is 'quickly' and 'correctly' impounded in prices and other quantities constituting the market model. Furthermore, $x^\top$ denotes again the transpose of $x$. For matrices $x$ and $y$ we write $x \cdot y$ for the matrix product of $x$ and $y$. Moreover, $\b1=(1, \dots, 1)^\top$ is a vector and we write $0$ for a zero matrix or vector, where the dimensions follow from the context.

Consider $m$ nonnegative, stocks with vector value process $S=\{S_t=(S^1_t, \dots$, $S^m_t)^\top$, $t \in [0,\infty)\}$, denominated in units of the domestic currency, which satisfies the stochastic differential equation (SDE)
\BE \label{e.2.1} \frac{dS_t}{S_t}=a_t dt + b_t \cdot dW_t,\EE
$t \in [0,\infty)$, $S^i_0 >0$ for $i = 1, \dots,m$. Note that all dividends are reinvested. The instantaneous expected return vector process $a=\{a_t=(a^1_t,\dots,a^m_t)^\top, t \in [0,\infty)\}$ and the volatility matrix process $b=\{b_t=[b^{j,k}_t]^{m,n}_{j,k=1},t\in (0,\infty]\}$ are assumed to be adapted and such that there exists a unique strong solution of the SDE (\ref{e.2.1}); see e.g. Section 7.7 in \citeN{PlatenHe10} for respective sufficient conditions.

 A strictly positive, self-financing, portfolio process $V^\pi$ is characterized by the weights or fractions of wealth $\pi_t=(\pi^1_t, \dots, \pi^m_t)^\top, t \in [0,\infty)$, invested in the stocks, together with its positive initial value $V^\pi_0>0$, where 
\BE \pi^\top_t \cdot \b1=1.\EE
 The portfolio value $V^\pi_t$ at time $t$ satisfies then the SDE
\BE \label{e.2.3}\frac{dV^\pi_t}{V^\pi_t}=\pi^\top_t \cdot \frac{dS_t}{S_t}=\pi^\top_t \cdot a_t dt+\pi^\top_t \cdot b_t \cdot dW_t\EE for $t \in [0,\infty)$. 

\subsection*{Growth Optimal Portfolio}

 We characterize the growth optimal portfolio (GP) by the following result, which follows directly from Theorem 3.1 in \citeN{FilipovicPl09}. There it has been shown that the GP is equivalent to the expected logarithmic utility maximizing portfolio, also called Kelly portfolio, discovered in \citeN{Kelly56} and studied in a stream of literature; see \shortciteN{MacLeanThZi11}. 

\begin{theorem}[Growth Optimal Portfolio Theorem]\label{t.b1}
 If a GP exists in the given continuous market, then the process $\pi^*$ of GP weights may not be unique. However, the GP value process $V^{\pi_*}=\{V^{\pi_*}_t, t \in [0,\infty)\}$ is unique for some fixed initial portfolio value, which we set to $V^{\pi_*}_0=1$, and the SDE of the GP is of the form
 \BE \label{e.4.1}\frac{dV^{\pi_*}_t}{V^{\pi_*}_t}=\lambda_t dt +\theta^\top_t \cdot (\theta_t dt+dW_t) \EE
 for $t \in [0,\infty)$. Here we set \BE \label{e.4.2} \theta_t=b^\top_t \cdot \pi^*_t,\EE
 with $\pi^*_t$ and $\lambda_t$ representing the components of the solution of the matrix equation
\BE \label{meq} \left( \begin{array}{cc} b_t b^\top_t & \b1  \\ \b1^\top & 0 \end{array} \right) \left( \begin{array}{c} \pi^*_t \\ \lambda_t  \end{array} \right)=\left( \begin{array}{c} a_t \\ 1 \end{array} \right) \EE
for all $t \in [0,\infty)$. A sufficient condition for the existence of a solution of (\ref{meq}) is the invertibility  of the covariance matrix $b_t \cdot b^\top_t$ for all $t \in [0,\infty)$. In the case when the risk-free asset is included in the investment universe $\lambda_t$ equals the risk-free rate and $\theta_t$ represents the vector of market prices of risk.
\end{theorem}

Assuming that the GP exists, it follows from the above Growth Optimal Portfolio Theorem (Theorem \ref{t.b1}) that the Lagrange multiplier $\lambda_t$ and the GP volatility vector $\theta_t$ are uniquely determined through $a_t$ and $b_t$. Moreover, by (\ref{e.4.2}) and (\ref{meq}) the vector of instantaneous expected returns has the form
\BE \label{e.5.2} a_t=\lambda_t \b1+b_t \cdot \theta_t.\EE
Consequently, for any self-financing portfolio $V^\pi_t$ the SDE (\ref{e.2.3}) takes the form
\BE \label{e.5.3}\frac{dV^\pi_t}{V^\pi_t}=\lambda_t dt+\pi^\top_t \cdot b_t (\theta_t dt+dW_t).\EE

To accurately identify the optimal strategy for the GP one would need accurate information about $a_t$ and $b_t$, which, as we argue in this paper, seems impossible to extract with sufficient precision to be useful in portfolio optimization for large equity markets. However, some reliable information is available through the hierarchical industrial and geographical groupings of stocks. As we show in this paper, this information is sufficient to approximate well enough the GP so that the following Efficient Market Property cannot be easily rejected.

\subsection*{Efficient Market Property}

In the benchmark pricing theory, see \citeN{PlatenHe10}, the GP of a given set of constituents is called benchmark. Any security or portfolio $V^\pi_t$ is called benchmarked when denominated in units of the benchmark. By applying the It\^{o} formula to the benchmarked portfolio value $\hat{V}^\pi_t=\frac{V^\pi_t}{V^{\pi_*}_t}$ we obtain from (\ref{e.5.3}) and (\ref{e.4.1}) for its return process $\hat{Q}^\pi=\{\hat{Q}^\pi_t, t\in [0,\infty)\}$ the SDE
\BE \label{e.5.4} d\hat{Q}^\pi_t=\frac{d \hat{V}^\pi_t}{\hat{V}^\pi_t}=\l(\pi^\top_t \cdot b_t -\theta^\top_t \r)dW_t\EE
for $t \in [0,\infty)$. The key observation is here that this SDE is driftless. Consequently, it emerges the following fundamental fact:  

\begin{theorem}\label{t.b2}
In a continuous market the instantaneous expected returns of by the GP benchmarked self-financing portfolios equal zero.
\end{theorem}
 
To make the Efficient Market Property testable we need to acknowledge the fact that we can only observe returns over a nonvanishing strictly positive time period. To cover the situation where we have a strictly positive time period over which we observe returns, we note that driftless benchmarked portfolio values form, so called, local martingales. By Fatou's lemma any nonnegative local martingale  is a supermartingale, which means that any nonnegative benchmarked portfolio $\hat{V}^\pi$ satisfies the inequality
\BE \label{b.10}\hat{V}^\pi_t \geq E_t(\hat{V}^\pi_{t+h})\EE for all $0 \leq t \leq t+h <\infty$; see e.g. \citeN{PlatenHe10}. Here $E_t$ denotes the conditional expectation under the real-world probability measure $P$ given the information $\mathcal{F}_t$ at time $t$. As indicated earlier, the supermartingale property (\ref{b.10}) holds generally in semimartingale markets, see \citeN{KaratzasKa07} and \citeN{DuPlaten16}.

By setting the benchmark equal to the GP one has only negative or zero (but never strictly positive) expected returns for nonnegative benchmarked securities, since by \eqref{b.10} we have
\BE \label{b.10p} E_t \left( \frac{\hat{V}^\pi_{t+h}-\hat{V}^\pi_t}{\hat{V}^\pi_t}\right) \leq 0\EE
for $0 \leq t \leq t+h < \infty$. This result together with Theorem \ref{t.b2} proofs the statement of Theorem \ref{t.emp}. 

\renewcommand{\theequation}{\mbox{C.\arabic{equation}}}
\renewcommand{\thefigure}{C.\arabic{figure}}
\renewcommand{\thetable}{C.\arabic{table}}
\renewcommand{\thesubsubsection}{C.\arabic{section}}
\renewcommand{\thetheorem}{C.\arabic{theorem}}
\renewcommand{\thecorollary}{C.\arabic{theorem}}
\renewcommand{\theassumption}{C.\arabic{theorem}}
\renewcommand{\thedefinition}{C.\arabic{definition}}

\section*{Appendix C: Hierarchical Diversification}\setcA \setcB

\subsection*{Hierarchical Grouping of Stocks}

To formulate the Diversification Theorem we make a few assumptions
that avoid technicalities in its formulation and proof. These assumptions can be significantly relaxed in an obvious manner. We assume that the stocks can be classified into hierarchical groupings
with a fixed number $H \in \{1, 2, \dots\}$ of hierarchical levels. For example, in the construction of the
HWI, shown in Figure \ref{f.00}, we chose $H = 4$. In
the asymptotics of the Diversification Theorem, we let a number $M \in \{2,3,\dots\}$ tend to infinity, whereas
$H$ remains fixed. We assume that in each group of the hierarchy we have at least $\underline{K}M$ and at most $\bar{K}M$ next lower level subgroups, with fixed integers $\underline{K}$ and $\bar{K}$, $0< \underline{K} \leq \bar{K}<\infty$. This means, for given $M$ we have at least $(\underline{K} M)^H$ and at most $(\bar{K}M)^H$ stocks in our investment
universe.

We denote by $W^{k_1} = \{W^{k_1}_t, t \in [0, \infty)\}$ the $k_1$-th independent, standard Brownian motion that drives primarily the $k_1$-th group on the highest level of the hierarchy, $k_1 \in \{1, 2, \dots, \bar{K}M \}$. Furthermore, for $k_1 , k_2 \in \{1,2, \dots,\bar{K}M \}$ we let
$W^{k_1,k_2} = \{W^{k_1,k_2}_t , t \in [0, \infty)\}$ denote the $(k_1, k_2)$-th independent standard Brownian motion
that primarily models the uncertainty driving the $k_2$-th group on the second
highest level in the $k_1$-th group of the highest level. In an analogous manner
we introduce independent standard Brownian motions for next lower level
groups until we reach at the lowest level the stocks. Here $W^{k_1,k_2,\dots, k_H}$ denotes the $(k_1,k_2,\dots,k_H)$-th independent standard Brownian motion that primarily drives the $k_H$-th stock in the $k_{H-1}$-th lowest level group, of the $k_{H-2}$-th second lowest level group, etc.  For the $j_H$-th benchmarked stock in the $j_{H-1}$-th lowest level group of the $j_{H-2}$-th second lowest level group, etc., we write $\hat{S}^j_t$, where $j=(j_1, j_2, \dots,j_H) \in \Gamma_{M}=(1,2,\dots,\bar{K}M)^H$. By using (\ref{e.2.1}) and (\ref{e.5.2}) we capture the hierarchical structure of the stock market dynamics for the $j$-th benchmarked stock price $\hat{S}^j_t$ by assuming the SDE 
\begin{eqnarray}\label{e.b10}
\frac{d \hat{S}^j_t}{\hat{S}^j_t}&=&\sum^{\bar{K}M}_{k_1=1}(\psi^{j,k_1}_t dW^{k_1}_t+\sum^{\bar{K}M}_{k_2=1}( \psi^{j,k_1,k_2}_tdW^{k_1,k_2}_t\\ \nonumber &+&\dots +\sum^{\bar{K}M}_{k_H=1} \psi^{j,k_1,k_2,\dots,k_H}_tdW^{k_1,k_2,\dots,k_H}_t )).
\end{eqnarray}
Note that by setting for some $j \in \Gamma_M$ the respective volatility coefficient to zero we can conveniently model groups that have less than $\bar{K}M$ next lower level subgroups. 

According to (\ref{e.b10}) the hierarchical groupings allow us to capture in each group the subgroups
that have exposure to similar industrial and geographical uncertainties. Note, the sources of uncertainty typical for members of a group are
only assumed to be of significance to other members of the group, which seems to be reasonable.

\subsection*{Diversification Theorem}

For given $M \in \{2,3,\dots\}$ the SDE for the return process $\hat{Q}^{\pi_M}_t$ of a given benchmarked portfolio $\hat{V}^{\pi_M}_t$, with fraction $\pi^j_{M,t}$ invested in the $j$th stock, $j \in \Gamma_M$, has then by (\ref{e.5.4}) the form
\begin{eqnarray}
 d\hat{Q}^{\pi_M}_t&=&\frac{d\hat{V}^{\pi_M}_t}{\hat{V}^{\pi_M}_t}=\sum_{j \in \Gamma_{M}} \pi^j_{M,t} \frac{d\hat{S}^j_t}{\hat{S}^j_t}\\
 &=& \sum^{\bar{K}M}_{k_1=1} \sum_{j \in \Gamma_{M}} \pi^j_{M,t} \psi^{j,k_1}_t dW^{k_1}_t\\
 &+& \sum^{\bar{K}M}_{k_1=1}\sum^{\bar{K}M}_{k_2=1}\sum_{j \in \Gamma_{M}} \pi^{j}_{M,t} \psi^{j,k_1,k_2}_t dW^{k_1,k_2}_t\\
 &+& \dots + \sum^{\bar{K}M}_{k_1=1}\sum^{\bar{K}M}_{k_2=1} \dots \sum^{\bar{K}M}_{k_H=1} \sum_{j \in \Gamma_{M}}  \pi^j_{M,t} \psi^{j,k_1,k_2, \dots, k_H}_tdW^{k_1,k_2,\dots,k_H}_t.
\end{eqnarray}
Note that the benchmarked GP equals trivially the constant one. Therefore, the
time derivative of the quadratic variation, see e.g. \citeN{PlatenHe10}, of its return process equals zero. Since we aim
to identify proxies of the GP, we identify these asymptotically as follows, where the time derivative of the quadratic variation of the return process of the benchmarked proxy vanishes as $M$ tends to infinity:

\begin{definition} \label{d.b4}
We call a sequence of benchmarked portfolios $(\hat{V}^{\pi_M})_{M \in \{2, 3,\dots\}}$, each with return process $\hat{Q}^{\pi_M}$, a sequence of approximate benchmarked GP processes if for all $\varepsilon >0$ and $t \in [0,\infty)$ the limit in probability \BE \lim_{M \rightarrow \infty} P(\frac{d[\hat{Q}^{\pi_M}_\cdot]_t}{dt} > \varepsilon)=0\EE holds.
\end{definition}

For diversification to be possible we need a condition which ensures that not all benchmarked stocks are driven to a significant extend by the same uncertainties. Therefore, we make the following rather reasonable assumption:

\begin{assumption} \label{a.b5} For given $k_1,k_2,\dots, k_h \in \{1,2,\dots,\bar{K}M\}$ we assume for all $M \in \{2,3,\dots\}$ and all $h \in \{1,2, \dots,H\}$ that
\BE \label{e.b20} \sum_{j \in \Gamma_{M}} |\psi^{j,k_1,k_2,\dots, k_h}_t| \leq (\bar{K}M)^{H-h}\sigma_t,\EE where the adapted stochastic process $\sigma=\{\sigma_t, t \geq 0\}$ satisfies the square integrability condition
\BE \label{e.b21} E((\sigma_t)^2)\leq \bar{\sigma}^2<\infty\EE for all $t \in [0,\infty)$.
\end{assumption}

The Assumption \ref{a.b5} covers an extremely wide range of realistic hierarchical market models. Note that the particular form of the volatilities $\psi^{j,\cdot}_t$ is here not relevant. What is limited by (\ref{e.b20}) is the sum of the absolute values of volatilities with respect to the same source of uncertainty. The above property secures some convergence towards the GP if the weights of the constituents vanish 'fast enough' for increasing $M$, as we specify in the Diversification Theorem below: 

\begin{theorem}[Diversification Theorem] \label{t.b6} A sequence of benchmarked portfolios $(\hat{V}^{\pi_M})_{M \in \{2, 3,\dots\}}$, is a sequence of benchmarked approximate GPs, if for each $M \in \{2, 3, \dots\}$ the maximum of the weights satisfies for some parameter $\xi \in [0,\frac{1}{2})$ and some constant $C \in (0,\infty)$ the relation
\BE \label{cond} \max_{j \in \Gamma_{M}} |\pi^j_{M,t}| \leq C  M^{\xi-H} \EE
for all $t \in [0,\infty)$.
\end{theorem}

Obviously, relation (\ref{cond}) is satisfied for the EWI and the HWI, since in these cases we have by (\ref{Nt}), (\ref{piEWI}) and (\ref{piHWI}) that $\pi^j_{M,t} \leq \underline{K}^{-H}M^{-H}$ with $C=\underline{K}^{-H}$ and $\xi=0$ in (\ref{cond}). Therefore, both indexes form sequences of approximate GPs. We will see below that the proof of the above Diversification Theorem exploits crucially the Efficient Market Property. 

The Diversification Theorem can also be intuitively interpreted as a consequence of a generalized version of the Law of Large Numbers for returns of benchmarked proxies of the GP. To see this in an illustrative situation, consider benchmarked constituents of the GP that have independent, square integrable returns. We know from the Efficient Market Property that these returns have asymptotically zero mean over vanishing time periods. When we form an equal-weighted index (EWI), the total return of the benchmarked EWI becomes the average of the independent returns. Thus, by the Law of Large Numbers this yields asymptotically zero returns for the benchmarked EWI for increasing number of constituents. Consequently, the benchmarked EWI equals in the limit the constant one. By multiplying the limiting benchmarked EWI, that is the constant value one, with the GP in domestic currency denomination we obtain the GP in domestic currency denomination. Thus the GP equals asymptotically the limit of the EWI in domestic currency. This illustration explains intuitively that, for increasing number of stocks, naive diversification approximates asymptotically the GP. The above Diversification Theorem identifies not only the EWI as a good proxy of the GP but also the HWI and many other well-diversified portfolios as good proxies. 

\subsubsection*{Proof of Theorem \ref{t.b6}} According to Definition \ref{d.b4} we need to study the time derivative of the quadratic variation of the return process $\hat{Q}^{\pi_M}$ of the benchmarked portfolio $\hat{V}^{\pi_M}$. The time derivative of the quadratic variation of the above return process $\hat{Q}^{\pi_M}$ equals 
\begin{eqnarray}\label{e.b30}\nonumber \frac{d[\hat{Q}^{\pi_M}_\cdot]_t}{dt}&=&\sum^{\bar{K}M}_{k_1=1}(\sum_{j \in \Gamma_{M}}\pi^j_{M,t} \psi^{j,k_1}_t)^2\\
\nonumber &+&\sum^{\bar{K}M}_{k_1=1}\sum^{\bar{K}M}_{k_2=1}(\sum_{j \in \Gamma_{M}}\pi^j_{M,t} \psi^{j,k_1,k_2}_t)^2\\
&+& \dots + \sum^{\bar{K}M}_{k_1=1}\sum^{\bar{K}M}_{k_2=1} \dots\sum^{\bar{K}M}_{k_H=1}(\sum_{j \in \Gamma_{M}}\pi^j_{M,t} \psi^{j,k_1,k_2,\dots,k_H}_t)^2
\end{eqnarray} for $t \in [0,\infty)$. 
By (\ref{e.b20}) we obtain from (\ref{e.b30}) that

\begin{eqnarray}
 \nonumber \frac{d[\hat{Q}^{\pi_M}_\cdot]_t}{dt} &\leq& (\max_{j \in \Gamma_{M}} |\pi^j_{M,t}|)^2 \Bigg{(}\sum^{\bar{K}M}_{k_1=1}(\sum_{j \in \Gamma_{M}}|\psi^{j,k_1}_t|)^2\\
\nonumber &+&\sum^{\bar{K}M}_{k_1=1}\sum^{\bar{K}M}_{k_2=1}(\sum_{j \in \Gamma_{M}}|\psi^{j,k_1,k_2}_t|)^2\\
\nonumber &+& \dots + \sum^{\bar{K}M}_{k_1=1}\sum^
{\bar{K}M}_{k_2=1} \dots\sum^{\bar{K}M}_{k_H=1}(\sum_{j \in \Gamma_{M}}| \psi^{j,k_1,k_2,\dots,k_H}_t|)^2 \Bigg{)}\\
\nonumber &\leq& (\max_{j \in \Gamma_{M}} |\pi^j_{M,t}|)^2  \sigma^2_t \sum^{H}_{h=1} (\bar{K}M)^h (\bar{K}M^{H-h})^2\\
\nonumber &\leq& (\max_{j \in \Gamma_{M}} |\pi^j_{M,t}|)^2  \sigma^2_t \bar{K}^{2H} \sum^{H}_{h=1} (M)^h (M^{H-h})^2\\
\nonumber &=& \sigma^2_t (\max_{j \in \Gamma_{M}} |\pi^j_{M,t}|)^2 \bar{K}^{2H} M^{2H-1} \sum^H_{h=1} \l( \frac{1}{M}\r)^{h-1}\\
\nonumber &\leq&  \sigma^2_t \bar{K}^{2H} (\max_{j \in \Gamma_{M}} |\pi^j_{M,t}|)^2 \frac{M^{2H-1}}{1-\frac{1}{M}}.
\end{eqnarray} The last estimate follows from the well-known formula for the limit of the sum of a geometric series. Thus, by using (\ref{cond}) we get
\begin{eqnarray} \label{e.b23} \frac{d[\hat{Q}^{\pi_M}_\cdot]_t}{dt}  &\leq& C^2 \sigma^2_t \bar{K}^{2H-h} M^{2(\xi-H)}M^{2H-1}(1-M^{-1})^{-1} \\ \nonumber &\leq& C^2 \sigma^2_t \bar{K}^{2H-h} M^{2\xi-1} (1-M^{-1})^{-1}\\
\nonumber &\leq& 2 C^2 \sigma^2_t \bar{K}^{2H-h} M^{2\xi-1}.\end{eqnarray}
Since $\xi \in [0,\frac{1}{2})$, it follows by the Markov inequality with (\ref{e.b21}) for each $\varepsilon>0$ and $t \in [0,T]$ that
\BE \label{end} \lim_{M \rightarrow \infty}P(\frac{d[\hat{Q}^{\pi_M}_\cdot]_t}{dt} > \varepsilon) \leq \lim_{M \rightarrow \infty } \frac{1}{\varepsilon}E(\frac{d[\hat{Q}^{\pi_M}_\cdot]}{dt}) \leq \frac{2C^2}{\varepsilon} \bar{\sigma}^2 \bar{K}^{2H} \lim_{M \rightarrow \infty} M^{2\xi-1}=0,\EE which proves the Diversification Theorem.
$\Square$

\renewcommand{\theequation}{\mbox{D.\arabic{equation}}}
\renewcommand{\thefigure}{D.\arabic{figure}}
\renewcommand{\thetable}{D.\arabic{table}}
\renewcommand{\thesubsubsection}{D.\arabic{section}}
\renewcommand{\thetheorem}{D.\arabic{theorem}}
\renewcommand{\thecorollary}{D.\arabic{theorem}}
\renewcommand{\theassumption}{D.\arabic{theorem}}
\renewcommand{\thedefinition}{D.\arabic{definition}}

\section*{Appendix D: A Stylized Hierarchical Market Model}\setcA \setcB
To give an idea why the HWI performs so well, even when in reality market prices of risk and other quantities are likely changing very fast and cannot be estimated, we describe in this section a rather realistic but stylized hierarchical stock market model that is covered by the Diversification Theorem. The key message is that it turns out that under this model the HWI coincides exactly with the GP. The somehow surprising observation is that the GP strategy does not request here any quantity as input that has to be estimated. The stylized model exploits the hierarchical stock market structure we assume. It then sets various stochastic quantities equal, which for economic reasons are most likely similar but have no chance to become estimated with any useful accuracy. The particular values of the randomly moving quantities turn out to be irrelevant for the optimal weights of the GP. We emphasize that under the following stylized model various key quantities can form fast fluctuating stochastic processes. 

Extremely difficult to estimate are quantities that characterize the mean-variance optimal wealth evolution of a particular company. The stylized model assumes that each company achieves a mean-variance optimal wealth evolution. Thus, by setting in our stylized model its total issued stock value equal to its wealth, the uncertainties of its economic activities become securitized in the stock. The level of risk aversion applied by the management of the company is a key quantity and cannot be easily dismissed when characterizing the company's mean-variance optimal wealth evolution. This risk aversion level most likely changes over time and most likely changes similarly for most companies. Therefore, the stylized model assumes for all companies a common risk aversion process, denoted by $\gamma=\{\gamma_t>0,t \geq 0\}$. 

The other crucial input that plays a significant role in characterizing the mean-variance optimal wealth evolution of a company is the vector of market price of risk processes for the various sources of uncertainty faced by the companies in the market. Also these processes most likely change over time and are difficult to estimate. All companies are exposed to similar prices for raw materials, energy, labor, etc. Consequently, common market price of risk processes can be assumed in the stylized model. Additionally, one could argue that capital most likely is flowing to those business opportunities facing slightly higher market prices of risk and is avoiding investments with lower market prices of risk. Therefore, market price of risk processes can be assumed in our stylized model to be equal but fluctuate over time. The common market price of risk process $\theta=\{\theta_t, t \geq 0\}$ is then assumed to denote the market price of risk for each of the independent sources of uncertainty in the stylized model.

 We then assume that in its mean-variance wealth optimization, the $j$-th company applies the risk aversion $\gamma_t$ at time $t$ to invest through its management activities the fraction $\frac{1}{\gamma_t}$ in its 'own' GP denoted by $S^{j,GP}_t$. This is the GP for the 'investment universe' determined by the business opportunities and activities of the $j$-th company. As is well-known, the company holds then the fraction $1-\frac{1}{\gamma_t}$ of its wealth in units of the risk-free asset; see e.g., \citeN{CampbellVi02} or Theorem 11.1.3 in \citeN{PlatenHe10}. The stochastic differential equation (SDE) for the stock price $S^j_t$ is then obtained in the form
\BE \label{e.dSj} \frac{dS^j_t}{S^j_t}= \frac{1}{\gamma_t} \frac{dS^{j,GP}_t}{S^{j,GP}_t}+\l(1-\frac{1}{\gamma_t}\r)r_t dt,\EE
where $r_t$ denotes the risk free rate.

To model the uncertainties faced by the $j$-th company, $j=(j_1,j_2,j_3,j_4)$ its wealth is driven by the specific uncertainty $W^{j_1,j_2, j_3,j_4}$, the uncertainty $W^{j_1,j_2,j_3}$  typical for the industrial grouping the company belongs to, the uncertainty $W^{j_1, j_2}$ specific for the country where the company is located and the uncertainty $W^{j_1}$ that is typical for the region of the company's country. Here $W^{j_1, j_2, j_3, j_4}$, $W^{j_1, j_2, j_3}$, $W^{j_1, j_2}$ and $W^{j_1}$ are independent standard Brownian motions. The SDE for the 'own' GP of the $j$-th company follows then from Theorem \ref{t.b1} in Appendix B in the form
\begin{eqnarray}\label{e.dSjGP}\frac{dS^{j, GP}_t}{S^{j,GP}_t}&=&r_t dt+\theta_t (\theta_t dt+dW^{j_1}_t)+\theta_t (\theta_t dt+dW^{j_1,j_2}_t)\\
\nonumber  &+&\theta_t (\theta_tdt +dW^{j_1,j_2,j_3}_t)+\theta_t (\theta_t dt+dW^{j_1,j_2,j_3,j_4}_t),\end{eqnarray}
for $t \geq 0$. We emphasize that the risk-free asset is here included in the 'investment universe' of the $j$-th company when forming its 'own' GP. When we form below the GP of the given set of stocks, we do not include the risk-free asset in this investment universe, since we aim in this case to identify the GP for the stocks only.

Following (\ref{e.dSj}) and (\ref{e.dSjGP}) the $j$-th cum-dividend stock value, with $j=(j_1,j_2,\\j_3,j_4)$, satisfies the SDE 
\begin{eqnarray}\label{e.c1}\frac{dS^j_t}{S^j_t}&=&r_t dt+\frac{1}{\gamma_t}\Big{(}\theta_t(\theta_t dt+dW^{j_1}_t)+\theta_t(\theta_t dt+dW^{j_1,j_2}_t))\\
\nonumber &+& \theta_t(\theta_t dt+dW^{j_1, j_2, j_3}_t)+\theta_t(\theta_t dt +dW^{j_1, j_2, j_3,j_4})\Big{)}.
\end{eqnarray} 

Based on the SDE (\ref{e.c1}) and the weights (\ref{piHWI}) the HWI under the stylized model has the return process 
\begin{eqnarray}
 \frac{dS^{HWI}_t}{S^{HWI}_t}&=& r_t dt+\frac{1}{\gamma_t} \frac{1}{M_t} \sum^{M_t}_{j_1=1} \Bigg{(}\theta_t(\theta_tdt+dW^{j_1}_t)\\
 \nonumber &+& \frac{1}{M^{j_1}_t} \sum^{M^{j_1}_t}_{j_2=1} \Bigg{(}\theta_t(\theta_t dt+dW^{j_1,j_2}_t)\\
 \nonumber &+&\frac{1}{M^{j_1,j_2}_t}\sum^{M^{j_1,j_2}_t}_{j_3=1}\Bigg{(}\theta_t(\theta_t dt+dW^{j_1,j_2,j_3}_t)\\
 \nonumber &+& \frac{1}{M^{j_1,j_2,j_3}_t}\sum^{M^{j_1,j_2,j_3}_t}_{j_4=1} \theta_t(\theta_t dt+dW^{j_1,j_2,j_3,j_4}_t)\Bigg{)}\Bigg{)}\Bigg{)}
\end{eqnarray} for $t \geq 0$ with $S^{HWI}_0>0$.

By application of the It\^{o} formula it is straightforward to show that the $j$-th stock, when denominated in units of the HWI, has zero drift. That is, we have for the benchmarked $j$-th stock $\hat{S}^j_t=\frac{S^j_t}{S^{HWI}_t}$ the SDE 
\begin{eqnarray} \frac{d\hat{S}^j_t}{\hat{S}^j_t}&=&\sum^{M_t}_{k_1=1} (\psi^{j,k_1}_t dW^{k_1}_t+\sum^{M^{k_1}_t}_{k_2=1}(\psi^{j,k_1,k_2}_t dW^{k_1,k_2}_t\\ \nonumber &+&\sum^{M^{k_1,k_2}_t}_{k_3=1}(\psi^{j,k_1,k_2,k_3}_t  dW^{k_1,k_2,k_3}_t
 + \sum^{M^{k_1,k_2,k_3}_t}_{k_4=1}\psi^{j, k_1,k_2,k_3,k_4}_t dW^{k_1,k_2,k_3,k_4}_t)))
\end{eqnarray}
with \BE \label{e.c4} \psi^{j, k_1,\dots, k_n}_t=\Bigg{\{}
\begin{array}{l}
\frac{1}{\gamma_t} \theta_t \Big{(}1-\frac{1}{M^{j_1}_t\dots \ M^{j_1,\dots,j_{n-1}}_t}\Big{)} \text{ for } k_i=j_i \text{ for all } i \in \{1,\dots,n\}\\
-\frac{1}{\gamma_t} \theta_t \frac{1}{M^{j_1}_t \dots \ M^{j_1,\dots,j_{n-1}}_t} \text{    \ \ \ \ \ \  otherwise}\\
\end{array}
\EE
for $n \in \{1,2,3,4\}$ with $M^{j_0}_t=M_t$, where we use again the previous notation. Since the SDE for the benchmarked $j$-th stock is driftless, it is a local martingale. This means, according to \citeN{FilipovicPl09} the HWI is the GP of the given stylized hierarchical market model when setting $S^{HWI}_0=S^{GP}_0=1$. This remarkable fact provides us with some extra intuition in the understanding why hierarchical diversification works well in practice despite the fact that the market price of risk processes and the risk aversion processes may fluctuate and are difficult to estimate. 
\bibliographystyle{chicago}

\begin{thebibliography}{}

\bibitem[\protect\citeauthoryear{Arnott, Hsu  \& Moore}{Arnott
  et~al.}{2005}]{ArnottHsMo05}
Arnott, R., J.~Hsu, \& P.~Moore (2005).
\newblock Fundamental indexation.
\newblock {\em Financial Analyst Journal\/}~{\bf 61}(2), 83--99.

\bibitem[\protect\citeauthoryear{Bai \& Ng}{Bai and Ng}{2002}]{BaiNg02}
Bai, J. \& S.~Ng (2002).
\newblock Determining the number of factors in approximate factor models.
\newblock {\em Econometrica\/}~{\bf 70}(1), 191--221.

\bibitem[\protect\citeauthoryear{Best \& Grauer}{Best and
  Grauer}{1991}]{BestGr91}
Best, M.~J. \& R.~R. Grauer (1991).
\newblock On the sensitivity of mean-variance-efficient portfolios to changes
  in asset means: {S}ome analytical and computational results.
\newblock {\em Review of Financial Studies\/}~{\bf 4}(2), 315--342.

\bibitem[\protect\citeauthoryear{Campbell \& Viceira}{Campbell and
  Viceira}{2002}]{CampbellVi02}
Campbell, J. \& L.~Viceira (2002).
\newblock {\em Strategic Asset Allocation, Portfolio Choice for Long Term
  Investors}.
\newblock Oxford, Oxford University Press.

\bibitem[\protect\citeauthoryear{Carhart}{Carhart}{1997}]{Carhart97}
Carhart, M. (1997).
\newblock On persistence in mutual fund performance.
\newblock {\em Journal of Finance\/}~{\bf 52}(1), 57--82.

\bibitem[\protect\citeauthoryear{Chopra \& Ziemba}{Chopra and
  Ziemba}{1993}]{ChopraZi93}
Chopra, V.~K. \& W.~Ziemba (1993).
\newblock The effect of error in means, variances and covariances in optimal
  portfolio choice.
\newblock {\em Journal of Portfolio Management\/}~{\bf 19}(1), 6--11.

\bibitem[\protect\citeauthoryear{Choueifaty \& Coignard}{Choueifaty and
  Coignard}{2008}]{ChoueifatyCo08}
Choueifaty, Y. \& Y.~Coignard (2008).
\newblock Towards maximum diversification.
\newblock {\em Journal of Portfolio Management\/}~{\bf 35}(1), 40--51.

\bibitem[\protect\citeauthoryear{Chow, Hsu, Kalesnik  \& Little}{Chow
  et~al.}{2011}]{ChowHsKaLi11}
Chow, T., J.~Hsu, V.~Kalesnik, \& B.~Little (2011).
\newblock A survey of alternative equity index strategies.
\newblock {\em Financial Analyst Journal\/}~{\bf 5}(67), 37--57.

\bibitem[\protect\citeauthoryear{Clarke, Silva  \& Thorley}{Clarke
  et~al.}{2011}]{ClarkeSiTh11}
Clarke, R., H.~Silva, \& S.~Thorley (2011).
\newblock Minimum-variance portfolio composition.
\newblock {\em Journal of Portfolio Management\/}~{\bf 37}(2), 31--45.

\bibitem[\protect\citeauthoryear{Davison \& Hinkley}{Davison and
  Hinkley}{2007}]{DavisonHi07}
Davison, A. \& D.~Hinkley (2007).
\newblock {\em Bootstrap Methods and their Applications}.
\newblock Cambridge Series in Statistical and Probabilistic Mathematics.
  Cambridge University Press.

\bibitem[\protect\citeauthoryear{DeMiguel, Garlappi  \& Uppal}{DeMiguel
  et~al.}{2009}]{DeMiguelGaUp09}
DeMiguel, V., L.~Garlappi, \& R.~Uppal (2009).
\newblock Optimal versus naive diversification: {H}ow inefficient is the $1/n$
  portfolio strategy?
\newblock {\em Review of Financial Studies\/}~{\bf 22}(5), 1915--1953.

\bibitem[\protect\citeauthoryear{Du \& Platen}{Du and
  Platen}{2016}]{DuPlaten16}
Du, K. \& E.~Platen (2016).
\newblock Benchmarked risk minimization.
\newblock {\em Mathematical Finance\/}~{\bf 26}(3), 617--637.

\bibitem[\protect\citeauthoryear{Fama}{Fama}{1970}]{Fama70}
Fama, E.~F. (1970).
\newblock Efficient capital markets: a review of theory and empirical work.
\newblock {\em The Journal of Finance\/}~{\bf 25}(2), 383--417.

\bibitem[\protect\citeauthoryear{Fama}{Fama}{1991}]{Fama91}
Fama, E.~F. (1991).
\newblock Efficient capital markets: {II}.
\newblock {\em The Journal of Finance\/}~{\bf 46}(5), 1575--1617.

\bibitem[\protect\citeauthoryear{Fama}{Fama}{1998}]{Fama98}
Fama, E.~F. (1998).
\newblock Market efficiency, long-term returns, and behavioral finance.
\newblock {\em Journal of Financial Economics\/}~{\bf 49}(3), 283--306.

\bibitem[\protect\citeauthoryear{Fama \& French}{Fama and
  French}{1992}]{FamaFr92}
Fama, E.~F. \& K.~R. French (1992).
\newblock The cross-selection of expected stock returns.
\newblock {\em The Journal of Finance\/}~{\bf XLVII}(2).

\bibitem[\protect\citeauthoryear{Fernholz}{Fernholz}{2002}]{Fernholz02}
Fernholz, E.~R. (2002).
\newblock {\em Stochastic Portfolio Theory}.
\newblock Springer.

\bibitem[\protect\citeauthoryear{Filipovi\'c \& Platen}{Filipovi\'c and
  Platen}{2009}]{FilipovicPl09}
Filipovi\'c, D. \& E.~Platen (2009).
\newblock Consistent market extensions under the benchmark approach.
\newblock {\em Mathematical Finance\/}~{\bf 19}(1), 41--52.

\bibitem[\protect\citeauthoryear{Gander, Leveau  \& Pfiffner}{Gander
  et~al.}{2013}]{GanderLePf13}
Gander, P., D.~Leveau, \& T.~Pfiffner (2013).
\newblock Diversification - a multi-facetted concept.
\newblock {\em 1741 Asset Management Research Note Series 1/2013\/}.

\bibitem[\protect\citeauthoryear{Harvey, Liu  \& Zhu}{Harvey
  et~al.}{2016}]{HarveyLiZh14}
Harvey, C., Y.~Liu, \& H.~Zhu (2016).
\newblock ... and the cross-section of expected returns.
\newblock {\em Review of Financial Studies\/}~{\bf 29}(1), 5--68.

\bibitem[\protect\citeauthoryear{Kan, Wang  \& Zhou}{Kan
  et~al.}{2016}]{KanWaZh16}
Kan, R., X.~Wang, \& G.~Zhou (2016).
\newblock On the value of portfolio optimization in the presence of estimation
  risk: The case with and without risk-free asset.
\newblock {\em Working paper. University of Toronto\/}.

\bibitem[\protect\citeauthoryear{Kan \& Zhou}{Kan and Zhou}{2007}]{KanZh07}
Kan, R. \& G.~Zhou (2007).
\newblock Optimal portfolio choice with parameter uncertainty.
\newblock {\em Journal of Financial and Quantitative Analysis\/}~{\bf 42},
  621--656.

\bibitem[\protect\citeauthoryear{Karatzas \& Kardaras}{Karatzas and
  Kardaras}{2007}]{KaratzasKa07}
Karatzas, I. \& C.~Kardaras (2007).
\newblock The numeraire portfolio in semimartingale financial models.
\newblock {\em Finance and Stochastics\/}~{\bf 11}(4), 447--493.

\bibitem[\protect\citeauthoryear{Karatzas \& Shreve}{Karatzas and
  Shreve}{1998}]{KaratzasSh98}
Karatzas, I. \& S.~E. Shreve (1998).
\newblock {\em Methods of Mathematical Finance}.
\newblock Springer.

\bibitem[\protect\citeauthoryear{Kardaras \& Platen}{Kardaras and
  Platen}{2010}]{KardarasPl10me}
Kardaras, C. \& E.~Platen (2010).
\newblock Minimizing the expected market time to reach a certain wealth level.
\newblock {\em SIAM Journal of Financial Mathematics\/}~{\bf 1}(1), 16--29.

\bibitem[\protect\citeauthoryear{Kelly}{Kelly}{1956}]{Kelly56}
Kelly, J.~R. (1956).
\newblock A new interpretation of information rate.
\newblock {\em Bell Systems Technology Journal\/}~{\bf 35}, 917--926.

\bibitem[\protect\citeauthoryear{Leote, Lu  \& Moulin}{Leote
  et~al.}{2012}]{LeoteLuMo12}
Leote, R., X.~Lu, \& P.~Moulin (2012).
\newblock Demystifying equity risk-based strategies: A simple alpha plus beta
  description.
\newblock {\em Journal of Portfolio Management\/}~{\bf 38}, 56--70.

\bibitem[\protect\citeauthoryear{Ludvigson \& Ng}{Ludvigson and
  Ng}{2007}]{LudvigsonNg07}
Ludvigson, S. \& S.~Ng (2007).
\newblock The empirical risk-return relation: A factor analysis approach.
\newblock {\em Journal of Financial Economics\/}~{\bf 83}(1), 171--222.

\bibitem[\protect\citeauthoryear{MacLean, Thorp  \& Ziemba}{MacLean
  et~al.}{2011}]{MacLeanThZi11}
MacLean, L., E.~Thorp, \& W.~Ziemba (2011).
\newblock {\em The Kelly Capital Growth Investment Criterion.}
\newblock World Scientific.

\bibitem[\protect\citeauthoryear{Maillard, Roncalli  \& Teiletche}{Maillard
  et~al.}{2010}]{MaillardRoTe10}
Maillard, S., T.~Roncalli, \& J.~Teiletche (2010).
\newblock The properties of equally weighted risk contribution portfolios.
\newblock {\em Journal of Portfolio Management\/}~{\bf 36}(4), 60--70.

\bibitem[\protect\citeauthoryear{Markowitz}{Markowitz}{1959}]{Markowitz59}
Markowitz, H. (1959).
\newblock {\em Portfolio Selection: Efficient Diversification of Investment}.
\newblock Wiley, New York.

\bibitem[\protect\citeauthoryear{Mode}{Mode}{1966}]{Mode66}
Mode, E. (1966).
\newblock {\em Elements of Probability and Statistics}.
\newblock Prentice-Hall, Englewood Cliffs, N.J.

\bibitem[\protect\citeauthoryear{Oderda}{Oderda}{2015}]{Oderda15}
Oderda, G. (2015).
\newblock Stochastic portfolio theory optimization and the origin of rule-based
  investing.
\newblock {\em Quantitative Finance\/}~{\bf 15}(8), 1259--1266.

\bibitem[\protect\citeauthoryear{Okhrin \& Schmid}{Okhrin and
  Schmid}{2006}]{OkhrinSc06}
Okhrin, Y. \& W.~Schmid (2006).
\newblock Distributional properties of portfolio weights.
\newblock {\em Journal of Economics\/}~{\bf 134}, 235--256.

\bibitem[\protect\citeauthoryear{Platen}{Platen}{2011}]{Platen11}
Platen, E. (2011).
\newblock A benchmark approach to investing and pricing. in:.
\newblock {\em MacLean, L.C. and Thorp, E. O. and Ziemba, W. (2011), The Kelly
  Capital Growth Investment Criterion. World Scientific\/}, 409--425.

\bibitem[\protect\citeauthoryear{Platen \& Heath}{Platen and
  Heath}{2010}]{PlatenHe10}
Platen, E. \& D.~Heath (2010).
\newblock {\em A Benchmark Approach to Quantitative Finance}.
\newblock Springer Finance. Springer.

\bibitem[\protect\citeauthoryear{Platen \& Rendek}{Platen and
  Rendek}{2012}]{PlatenRe12an}
Platen, E. \& R.~Rendek (2012).
\newblock Approximating the num\'{e}raire portfolio by naive diversification.
\newblock {\em Journal of Asset Management\/}~{\bf 13}(1), 34--50.

\bibitem[\protect\citeauthoryear{Plyakha, Uppal  \& Vilkov}{Plyakha
  et~al.}{2014}]{PlyakhaUpVi14}
Plyakha, Y., R.~Uppal, \& G.~Vilkov (2014).
\newblock Equal or value weighting? {I}mplications for asset pricing tests.
\newblock Technical report, SSRN 1787045.

\bibitem[\protect\citeauthoryear{Reuters}{Reuters}{2008}]{Reuters08}
Reuters (2008).
\newblock Datastream global equity indices. {U}ser {G}uide. {I}ssue 5.
\newblock {\em Thomson Reuters\/}.

\bibitem[\protect\citeauthoryear{Rosenberg \& Marathe}{Rosenberg and
  Marathe}{1976}]{RosenbergMa76}
Rosenberg, B. \& V.~Marathe (1976).
\newblock Common factors in security returns: Microeconomic determinants and
  macroeconomic correlates.
\newblock {\em University of California Institute of Business and Economic
  Research, Research Program in Finance, Working paper No. 44\/}.

\bibitem[\protect\citeauthoryear{Sharpe}{Sharpe}{1964}]{Sharpe64}
Sharpe, W.~F. (1964).
\newblock Capital asset prices: {A} theory of market equilibrium under
  conditions of risk.
\newblock {\em Journal of Finance\/}~{\bf 19}, 425--442.

\bibitem[\protect\citeauthoryear{Van~Dijk}{Van~Dijk}{2011}]{VanDijk11}
Van~Dijk, M. (2011).
\newblock Is size dead? {A} review of the size effect in equity returns.
\newblock {\em Journal of Banking and Finance\/}~{\bf 35}(12), 3263--3274.

\end{thebibliography}

 \end{document}